\documentclass[reprint,aps,pre]{revtex4-1}
\usepackage{natbib}
\usepackage{graphicx}
\usepackage{amsmath,amssymb}
\usepackage[toc,page]{appendix}
\usepackage{color}
\usepackage{multirow}

\newcommand{\nocontentsline}[3]{}
\newcommand{\tocless}[2]{\bgroup\let\addcontentsline=\nocontentsline#1{#2}\egroup}

\definecolor{antiquefuchsia}{rgb}{0.57, 0.36, 0.51}

\begin{document}
\title{Balancing selfishness and norm conformity can explain human behavior in 
large-scale Prisoner's Dilemma games and can poise human groups near 
criticality}
\author{John Realpe-G{\'o}mez}\email{john.realpe@gmail.com}
\affiliation{Quantum Artificial Intelligence Laboratory, NASA Ames Research 
Center, Moffett Field, CA 94035, USA}
\affiliation{Instituto de Matem{\'a}ticas Aplicadas, Universidad de Cartagena, 
Cartagena de Indias, Bol{\'i}var 13001, Colombia}
\affiliation{SGT Inc., 7701 Greenbelt Rd., Suite 400, Greenbelt, MD 20770, USA}
\author{Giulia Andrighetto}
\affiliation{Institute of Cognitive Sciences and Technologies, National 
Research Council, Rome, 00185 Italy}
\affiliation{M\" alardalens University, Sweeden}
\affiliation{Institute for Futures Studies, Sweeden}
\author{Luis Gustavo Nardin}
\affiliation{Institute of Cognitive Sciences and Technologies, National 
Research Council, Rome, 00185 Italy}
\affiliation{Brandenburg University of Technology, Cottbus, 03046 Germany}
\author{Javier Antonio Montoya}
\affiliation{Instituto de Matem{\'a}ticas Aplicadas, Universidad de Cartagena, 
Cartagena de Indias, Bol{\'i}var 13001, Colombia}
\affiliation{Associates Program, The Abdus Salam International Centre for 
Theoretical Physics, Strada Costiera 11, 34151, Trieste, Italy}

\begin{abstract}

Cooperation is central to the success of human societies as it is crucial for 
overcoming some of the most pressing social challenges of our time. Yet how 
human cooperation is achieved and may persist is still a main puzzle in the 
social and biological sciences. Recently, scholars have recognized the 
importance of social norms as solutions to major local and large-scale 
collective action problems, from the management of water resources to the 
reduction of smoking in public places to the change in fertility practices. Yet 
a well-founded model of the effect of social norms on human cooperation is 
still lacking. Using statistical physics techniques and integrating findings 
from cognitive and behavioral sciences, we present an analytically-tractable 
model in which individuals base their decisions to cooperate both on the 
economic rewards they obtain and on the degree to which their action comply 
with social norms. Results from this parsimonious model are in  agreement with 
what has been observed in recent large-scale experiments with humans. We also 
find the phase diagram of the model and show that the experimental human group 
is poised near a critical point, a regime where recent work suggests living 
systems respond to changing external conditions in an efficient and coordinated 
manner.

\end{abstract}

\maketitle

\tocless\section{\label{s:intro}Introduction}
Cooperation is crucial to human social life, from friendship and professional 
relationships, to political participation and global level issues, like 
ecological conservation and international relations. Yet cooperation is often 
individually costly, making it inherently fragile. Many scholars have then been 
concentrated on understanding how to sustain it.

Mechanisms such as reputation~\cite{Milinski-Nature-2002}, communication and 
sanction~\cite{citeulike:1077227}, as well as social identity-related 
factors~\cite{yamagishi_group_2000} have been found to play a key role in 
promoting human cooperative behavior.

Recent solid empirical and field work evidence is mounting to suggest that 
social norms are successful in the provision and maintenance of cooperation in 
everyday life 
~\cite{ostrom2005,bowlesGints2011,Janssen2010,fehrFischbacher2004,Nyborg42}.
 Social norms are informal rules that prescribe what individuals ought or ought 
not to do and are typically enforced through informal sanctions, like 
ostracism, negative gossip, shame or 
disapproval~\cite{bicchieri2006,conteEtAl2013,citeulike:3420759}. 
They sustain behavior through shared beliefs and reciprocal expectations 
regarding the appropriate actions to perform in specific circumstances. 
Indispensable to social life, they are referred to as the 
`cement'~\cite{citeulike:3420759} or the `grammar' of 
society~\cite{bicchieri2006}.

Despite their importance, a rigorous and well-grounded model of how social 
norms affect human cooperative behavior is still lacking (see 
Sec.~\ref{s:assumptions}). Using statistical physics techniques and consistent 
with findings from the cognitive and behavioral  
sciences~\cite{chudekHenrich2011,bicchieri2006,bowlesGints2011,JEEA:JEEA12006}, 
we develop here an analytically-tractable model in which the decision makers' 
utility is based on a balancing between the material rewards they obtain and on 
the degree to which their action is in agreement with social norms. We  
explicitly incorporate the human ability to be sensitive to social 
norms---their so called norm psychology~\cite{chudekHenrich2011}---into the 
\textit{Experience Weighted Attraction} (EWA)~\cite{camererHo1999} framework.{ 
EWA is} a modeling approach that combines both reinforcement 
learning~\cite{suttonBarto1998} and belief learning~\cite{feltovich2012} that 
has been extensively explored in the field of behavioral economics and rather 
successful in explaining the interactive learning of humans in 
games~\cite{HoEtAl2007,citeulike:10311363}.

Results from our cognitively inspired model are in agreement with observations 
from recent large-scale experiments with humans (625 subjects) playing 
simultaneously large-scale Prisoner's Dilemma (PD) 
games~\cite{garcia-lazaroEtAl2012}.

The model \textit{quantitatively} reproduces both the global cooperation level 
(i.e., a decay from an initial value of 60\% to around 35\%) and the final 
distribution of agents according to their probability of cooperation. To the 
best of our knowledge, this is the first work that quantitatively reproduces 
both characteristics. The best attempts we know of are reported in 
Refs.~\cite{ciminiSanchez2014,ciminiSanchez2015,Vilone-PRE-2014,ezaki2016reinforcement,horitaEtAl2017} but, except for Ref.~\cite{horitaEtAl2017}, the focus 
of those works are on a \textit{qualitative} rather than quantitative 
understanding. The experiments studied in Ref.~\cite{horitaEtAl2017} have some differences with the type of experiments we analyze here, rendering a careful comparison more difficult (see Sec.~\ref{s:assumptions} for a discussion of the main differences). Furthermore, the models presented in those  works are not 
necessarily based on empirically-grounded cognitively-motivated assumptions, as 
the one we introduce here.

Our model is also parsimonious enough to allow for a detailed 
characterization of its long-term dynamics. We identify three parameter regimes 
where the system can be mono-stable, bi-stable, or remain out of equilibrium. 
Such regimes are separated by surfaces that terminate on a line of critical 
points, where it is well-known that systems can develop long range correlations 
and become highly responsive to external 
stimuli~\cite{munoz2017colloquium,hidalgoEtAl2014,Mora-StatPhys-2011,gelblumEtAl2015,Tkacik,Chate-Physics-2014,bialekEtAl2014,attanasiEtAl2014,Krotov2014,Nykter12022008,Mora2009,Beggs2008}.

Our findings suggest that groups of individuals who base their choice to 
cooperate on a balancing between selfishness and compliance with social norms 
poise near a critical point, where their capacity to respond efficiently to 
changing and widely diverse external conditions can be 
enhanced~\cite{hidalgoEtAl2014}. To the best of our knowledge, this is the 
first experimental evidence that human cooperative groups may operate near 
criticality (see e.g., Sec. IV of the very recent review in 
Ref.~\cite{munoz2017colloquium} for a detailed description of relevant works). 
This result hence points to an unexplored feature of human cooperation that may 
suggest a way in which social norms, besides promoting cooperation, can also 
enhance the ability of human groups to adapt to external variability. 
Similar results have been found in experiments with ants~\cite{gelblumEtAl2015}.

This work is outlined as follows. In Sec.~\ref{s:assumptions} we discuss 
previous research and provide an overview of the different components and 
assumptions of our agent-based model. In Sec.~\ref{s:learning} we describe the 
learning component of the model. In Sec.~\ref{s:ewan} we describe how agents 
make decisions by balancing individual and normative considerations. In 
Sec.~\ref{s:network} we make use of two further assumptions consistent with 
experiments, i.e., slow adaptation and absence of network reciprocity, to turn 
the stochastic agent-based model on networks presented in Sec.~\ref{s:ewan} 
into a four-parameter deterministic model of a single representative agent. In 
Sec.~\ref{s:dynamic} we determine the phase diagram of the effective 
single-agent model obtained in Sec.~\ref{s:network} and show that the model can 
display critical phenomena. In Sec.~\ref{s:experiments} we extract the 
parameters of the effective single-agent model from experimental data and show 
that human groups playing in the experiments are posed near criticality. In 
Sec.~\ref{s:conclusion} we present the conclusions of the work. In the 
Appendices we present further technical details.

\

\tocless\section{\label{s:assumptions}Previous work and model overview}
While there is a large number of literature on physics-based models of human 
cooperation (see e.g., Ref.~\cite{perc2017statistical} for a recent review), 
most of these models are theoretical works that do not take into account 
experimental evidence. Already about a decade ago a relevant review 
article~\cite{castellano2009statistical} noticed that the `contribution of 
physicists in establishing social dynamics as a sound discipline grounded on 
empirical evidence has been so far insufficient'. In a  recent 
`mini-review'~\cite{sanchez2017physics}, one of the leading researchers in the 
field remarked that even though `there are many relevant experimental results on cooperation on structured populations published in widely read journals while, unfortunately, many models are introduced in the literature without taking into account [such experimental] facts'.  Some of the most relevant 
experimental findings, as summarized by Sanchez~\cite{sanchez2017physics}, are:
(i) lattices or networks do not support cooperation; 
(ii) people display \textit{Moody Conditional Cooperation} (MCC), i.e., when 
deciding to cooperate individuals are responsive to the behavior of others, but 
only if they have cooperated themselves;
(iii) people do not take into account the earnings of their neighbors; and
(iv) cooperation can be sustained in dynamic networks.

Indeed, as pointed out in the Introduction, we have identified only few 
references~\cite{ciminiSanchez2014,ciminiSanchez2015,Vilone-PRE-2014,ezaki2016reinforcement,horitaEtAl2017}
 that have attempted to build empirically grounded models to explain the type 
of experiments we analyze here. However, except for Ref.~\cite{horitaEtAl2017}, 
the focus of those works was on obtaining a \textit{qualitative} understanding 
of the phenomena observed in this type of experiments.

In contrast, our work, as well as that by Horita et al.~\cite{horitaEtAl2017}, are {\it quantitative} studies. Horita et al.~\cite{horitaEtAl2017}  
compare the explanatory power of models of conditional 
cooperation~\cite{fischbacherEtAl2001,kesserVanWinden2000} and their moody 
variant (MCC)~\cite{garcia-lazaroEtAl2012} to reinforcement learning models in 
explaining cooperation under multiplayer social dilemma games. They fit these 
models to empirical data obtained from behavioral experiments, namely Prisoner's 
Dilemma and Public Goods Games. However, because their experiments have some 
differences with the type of experiments we analyze here, rendering a careful 
comparison is more difficult. For instance, while we analyze experiments with 625 
subjects interacting on a network during 52 rounds, Horita et al. study 
experiments where 100 individuals interact during 20 rounds either within fixed 
groups of four people each or with groups of four individuals chosen at random. The authors 
then aggregate the decisions made by individuals of all groups during all the 
rounds into a single dataset (see e.g. Eq. (13) in Ref.~\cite{horitaEtAl2017}). 
It is not clear to us whether relevant dynamical information is not lost in 
this aggregation process. In contrast, we extract our model parameters from 
relevant statistical features of individual large-scale experiments using 
techniques that explicitly acknowledge the dynamical nature of our model (see 
e.g. Eq.~\eqref{e:PbayesJoint} in Appendix~\ref{s:inference}).

Horita et al.~\cite{horitaEtAl2017} provide evidence that (model-free) reinforcement learning algorithms where agents have no access to information about decisions made by their neighbors can account for the observed human behavior roughly as accurately as algorithms where agents can directly encode the MCC rule. This result is particularly evident in those treatments in which subjects interact with different people at every stage, i.e., where norms and expectations about the actions of others are more difficult to emerge. 
This finding is consistent with evidence from the cognitive and behavioral sciences that inspired our model ~\cite{bowlesGints2011,citeulike:6670342} showing that although reinforcement learning plays an important role in governing human behavior, when involved in repeated and long-term interactions with the same people, individuals' choices are not independent from other people's behavior, but highly conditional on what they believe others will do. 

In the experiments analyzed by Horita et al. and those we analyze here the information about neighbor's decisions necessary to compute the normative reasoning (see Sec.~\ref{s:dN}) can in principle be extracted from the material payoffs. So, it is not unreasonable to expect that in these experiments subjects can indirectly infer normative information from material payoffs only, as suggested by Horita et al. A possible way to resolve this ambiguity in the future could be to design experiments where this peculiar situation does not hold.

On the other hand, theoretical and empirical evidence suggests that human 
strategic behavior is based not only on model-free reinforcement learning but 
also on model-based reinforcement learning (i.e., belief learning)~\cite{Lee-Cell-2016,camererHo1999,Glimcher-book-2013}. These two types of algorithms are related to habits that subjects acquired from past experiences and goals that they expect to achieve in the future, respectively. In contrast to Horita et al., our EWA-inspired model is a hybrid between these two learning algorithms. Although we assume an equal weight for both model-free and model-based reinforcement to simplify the analysis, the EWA component of our model can be easily generalized to incorporate the desired weight to each of these two algorithms. In future studies, this could be used to investigate which of the two approximations is more accurate, i.e., assuming all weight on model-free reinforcement learning as in Ref.~\cite{horitaEtAl2017} or assuming equal weight for both model-free and model-based reinforcement learning as we do here.

Additionally, the model we present here encodes empirically-grounded cognitive assumptions, as summarized in Table~\ref{t:assumptions}.

\begin{table*}\centering
	\begin{tabular}{l | l | l | l | l}
		\hline
		\multicolumn{2}{c|}{\bf Assumption} & \multicolumn{1}{c|}{\bf Description } & \multicolumn{1}{c|}{\bf 
		Representation} & \multicolumn{1}{c}{\bf Reference}\\
		\hline
		\multirow{8}{*}{\rotatebox{90}{\centering \textbf{1\textsuperscript{st} 
		block}}} & Bounded rationality						& Agents do not always play the 
		optimal strategy					& $\beta$ in Eq.~\eqref{eq:1} & 
		\cite{camererHo1999,Galla-PNAS-2013, Sato-PNAS}\\
		& \multirow{2}{*}{Belief learning}								& Agents learn from what 
		could have {\it potentially}			& \multirow{2}{*}{Eq.~\eqref{eq:d}} & 
		\multirow{2}{*}{\cite{camererHo1999,Galla-PNAS-2013, Sato-PNAS}} \\
		& & happened & & \\
		& Reinforcement learning				& Agents learn from what {\it actually} 
		happened					& Eq.~\eqref{eq:d} & \cite{camererHo1999,Galla-PNAS-2013, 
		Sato-PNAS}  \\
		& Memory decay									& Agents give more relevance to recent 
		events							& $\alpha$ in Eq.~\eqref{eq:d} & 
		\cite{camererHo1999,Galla-PNAS-2013, Sato-PNAS}\\
		& \multirow{2}{*}{Selfishness}	& Agents base their decisions on 
		self-regarding									& \multirow{2}{*}{$\Delta I_C,\, \Delta 
		I_D$, Eqs.~\eqref{eq:utility} and \eqref{e:Idrive}} & 
		\multirow{2}{*}{\cite{camererHo1999,Galla-PNAS-2013, Sato-PNAS}} \\
		& & considerations & & \\
		\hline
		\multirow{8}{*}{\rotatebox{90}{\centering \textbf{2\textsuperscript{nd} 
		block}}} & Norm conformity:							& Agents base their decisions {\it also} 
		on social norms				& $h$ in Eqs.~\eqref{eq:utility} and 
		\eqref{e:Ndrive} & \cite{andrighettoEtAl2013,cialdiniGoldstein2004} \\
		& \multirow{2}{*}{- Self-consistency}	& Agents are consistent with own 
		beliefs and				& \multirow{2}{*}{$w_C$ in Eq.~\eqref{e:Ndrive}} & 
		\multirow{2}{*}{\cite{festinger1957,abelsonBernstein1963,ayalGino2011}} \\
		& & self-ascribed norms & & \\
		& \multirow{2}{*}{- Social influence}		& Norm compliance increases with 
		the number of								& \multirow{2}{*}{$w_O$ in Eq.~\eqref{e:Ndrive}} & 
		\multirow{2}{*}{\cite{traylsenEtAl2010,fischbacherEtAl2001}} \\
		& & compliant peers & & \\
		& \multirow{2}{*}{- Moody conditional coop.}	& Social influence is 
		stronger if aligned with						& \multirow{2}{*}{$w_I$ in 
		Eq.~\eqref{e:Ndrive}} & \multirow{2}{*}{\cite{grujicEtAl2010}} \\
		& & self-consistency & & \\
		\hline
		\multirow{4}{*}{\rotatebox{90}{\centering \textbf{3\textsuperscript{rd} 
		block}}} & \multirow{2}{*}{Slow adaptation}							& Adaptation 
		happens over several individual							& 
		\multirow{2}{*}{Eqs.~\eqref{e:x-update-det-nonMF} and \eqref{e:dut}} & 
		\multirow{2}{*}{\cite{garcia-lazaroEtAl2012,grujicEtAl2014}} \\
		& & strategic choices & & \\
		& \multirow{2}{*}{No network reciprocity}			& Interaction structure does 
		not significantly						& \multirow{2}{*}{Eqs.~\eqref{eq:dynamic} and 
		\eqref{eq:mfUx}} & 
		\multirow{2}{*}{\cite{Sanchez-2015,garcia-lazaroEtAl2012,gutierrez-roigEtAl2014,traylsenEtAl2010,grujicEtAl2014}}
		 \\
		& & influence behavior & & \\
		\hline
	\end{tabular}
	\caption{Summary of assumptions underlying model presented here. The first 
	block of assumptions corresponds to the Experience Weighted Attraction (EWA) 
	model introduced in Ref.~\cite{camererHo1999}, restricted to the special case 
	discussed in the Supplementary Information of Ref.~\cite{Galla-PNAS-2013} 
	[see Sec. I there; cf. Ref.~\cite{Sato-PNAS}]. More specifically: (i) while 
	EWA allows for belief and reinforcement learning to have different weights, 
	here they have the same weight [see Eq.~\eqref{eq:d}]; (ii) while EWA allows 
	for the interpolation between average and cumulative reinforcement learning, 
	here the focus is exclusively on cumulative reinforcement learning [see 
	Eq.~\eqref{eq:d}]. These leads to a model characterized by the drive [see 
	Eq.~\eqref{eq:d}] and two parameters: (i) parameter $\alpha$ that captures 
	the exponential decrease of the relevance of past events; if $\alpha = 1$ 
	agents only remember what happened in the previous round, while if $\alpha = 
	0$ agents have cumulative information of the full history of play; (ii) 
	parameter $\beta$ that captures the success of agents in choosing the optimal 
	strategy; if $\beta \gg 1$ agents usually choose the optimal strategy, while 
	if $\beta = 0$ agents choose strategies at random. To the best of our 
	knowledge, the EWA model is based exclusively on self-regarding 
	considerations. The second block of assumptions extend the EWA model to 
	include norm-based considerations. The importance agents give to normative 
	considerations is characterized by parameter $h$; if $h = 0$ (if $h \gg 1$) 
	only individual (normative) considerations matter. The normative component 
	implements three processes characterized by parameters $w_C$, $w_O$ and 
	$w_I$. The more the norm is perceived as salient, i.e. relevant, by the agent 
	the higher its impact on the agent's decision. These parameters determine how 
	the norm salience is updated. Parameter $w_C$, however, can be absorbed in 
	parameter $h$, so we take $w_C = 1$. These two blocks of assumptions lead to 
	a stochastic agent-based model where interactions take place on a given 
	network. The third block of assumptions transform the model into a 
	deterministic analytically-tractable model of a single representative agent 
	characterized by parameter $\alpha$ and three \textit{effective} parameters 
	[see Eqs.~\eqref{eq:dynamic}-\eqref{e:y0}]; these three effective parameters 
	fully specify the long-term dynamics of the model (see 
	Sec.~\ref{s:ewan}).}\label{t:assumptions}
\end{table*}

The first block of assumptions in Table~\ref{t:assumptions} are specific to the 
EWA learning algorithm (see Sec.~\ref{s:learning}). While 
Refs.~\cite{ciminiSanchez2014,ciminiSanchez2015,Vilone-PRE-2014} implement a 
heuristic evolutionary dynamics, none actually implements the EWA learning 
dynamics~\cite{camererHo1999}, which is based on empirically sounder cognitive 
assumptions. Indeed, in Ref.~\cite{ciminiSanchez2014} authors recognize that 
`the original formulation of EWA cannot be trivially generalized to our MCC 
scenario' and attempt to reproduce key features of the EWA updating by a linear 
combination of belief and reinforcement learning (see Supplementary Information 
of Ref.~\cite{ciminiSanchez2014} under the section titled `SI EWA'). EWA, however, is known to be a better model than such a mixture (see e.g., item $3$ in page 323 of Ref.~\cite{camererHo1998}). Furthermore, in EWA agents learn solely from what they earned or could have earned, in agreement with experimental finding (iii) above.

The second block of assumptions in Table~\ref{t:assumptions} are specific to 
the normative component. These assumptions rely on theoretical and empirical 
studies showing that human decisions are not only driven by selfish 
considerations but also influenced by social norms (i.e., informal social rules 
prescribing what individuals ought or ought not to 
do~\cite{chudekHenrich2011,JEEA:JEEA12006,andrighettoEtAl2013}). Moreover, 
those postulations are also aimed to account for the fact that the more 
salient---i.e., relevant---the norm is perceived to be, the stronger its impact on 
the individual's motivation to comply with it. Vilone et 
al.~\cite{Vilone-PRE-2014} point out that the interplay of social and 
strategic motivations in human interactions is a largely unexplored topic in 
collective social phenomena. They implement a heuristic algorithm where, at 
each iteration, agents choose with a certain probability either to update their 
strategy by imitating a neighbor picked at random or to update their strategy 
based on strategic considerations. In addition to being heuristic, not 
necessarily based on empirical evidence, in the strategic component of their 
update rule  agents take into account the earning of their neighbors in 
contrast with the experimental finding (iii) above.

Our model, in addition to respecting experimental finding (iii), incorporates empirically grounded normative assumptions into the EWA framework while still conserving its general structure (see Sec.~\ref{s:ewan}). Apart from being affected by the expectations and actions of their peers, individuals' decision to cooperate depends also on their `mood'. Consistently with experimental finding (ii) above, when deciding to cooperate agents  are responsive to the behavior of others, but only if they have cooperated themselves.

The first two blocks of assumptions lead to a stochastic agent-based model in which agents interact on a given (static) network and  balance individual and normative considerations in their decision-making. It should not be difficult to extend this model to incorporate dynamical networks that can also take into account experimental finding (iv) above. However, we here restrict our analysis to static networks, which allows for further simplifications.

Finally, the third block of assumptions is also consistent with experimental evidence. Indeed, the nearly linear trend that usually characterizes the MCC rule in the type of experiments we analyze (see e.g., Figs. 3A and 3B in Ref.~\cite{garcia-lazaroEtAl2012}) is consistent with a relatively large randomness in agents' strategic choices (i.e., parameter $\beta \ll 1$; see Table~\ref{t:assumptions}). This implies that the time scale associated to individual strategic choices is smaller than the time scale on which adaptation happens, i.e., adaptation can be assumed slow; this assumption allows to turn the stochastic agent-based model into a deterministic one (see Sec.~\ref{s:network}). The second assumption in this block exploits experimental finding (i) to turn the resulting deterministic model of agents interacting on a (static) network into an effective four-parameter model of a single representative agent (see Sec.~\ref{s:network}). This four-parameter model is parsimonious enough to allow for the analytical determination of its long-term dynamics (see Sec.~\ref{s:dynamic}). We emphasize once again that this effective model is restricted to the study of interactions on static networks.

A remark is in order: While the reinforcement learning algorithms studied 
by Horita et al.~\cite{horitaEtAl2017} conserve the identity of the 
individuals, our mean field model just described is based solely on a single 
representative agent. While our single-agent model depends on four parameters, 
the two models studied by Horita et al. depend only on two or three parameters. 
However, our mean field four-parameter model is parsimonious enough to allow 
for the analytical characterization of its different dynamical regimes. 
Avoiding the adiabatic and mean field approximations described above, as well 
as the equal weights between model-free and model-based reinforcement learning, 
we could turn our model into a stochastic agent-based model that includes as 
special case the two-parameter model studied by Horita et al. In the future, 
such more general model could be used along with model selection techniques to 
better compare ours with the work of Horita et al.

\

\tocless\section{\label{s:learning}Learning algorithm}
Here we describe the learning component of the EWA model, which incorporates 
the first block of assumptions in Table~\ref{t:assumptions}. In the next 
section, we discuss how to extend this model to include normative 
considerations in agents' decision-making.

Theoretical and empirical evidence shows that human social strategic behavior 
is based on a combination of model-free and model-based reinforcement learning 
algorithms~\cite{Lee-Cell-2016,camererHo1999,Glimcher-book-2013}. These models 
are related to habits that subjects acquired from past experiences and goals 
that they expect to achieve in the future, respectively. Under some 
circumstances~\cite{Galla-PNAS-2013,Sato-PNAS}, carefully described in the 
first section of the supplementary information of Ref.~\cite{Galla-PNAS-2013}, 
this can be captured by a simplified form of the EWA model 
~\cite{camererHo1999}. EWA is a modeling approach that combines both 
reinforcement learning~\cite{suttonBarto1998} and belief 
learning~\cite{feltovich2012}. The former refers to reinforcing actions based 
on agents' past performance, and the latter refers to reasoning about how 
actions that have not been chosen would have performed. One of the key insights 
provided by EWA is that belief learning can also be understood as a process in 
which actions are reinforced by forgone payoffs. In this sense, EWA is a 
combination of model-free and model-based reinforcement 
learning~\cite{Lee-2013}. The simplified EWA 
model~\cite{Galla-PNAS-2013,Sato-PNAS} which we are interested in here can be 
described by the  equations

\begin{eqnarray}
\label{eq:1}x_{i}(t+1) = \frac{1}{1 + e^{-\beta D_{i}(t+1)}}, \\
\label{eq:d}D_{i}(t+1) = \left( 1 - \alpha \right) D_{i}(t) + \Delta U_{i}(t).
\end{eqnarray}
Here $x_{i}(t+1)$ and $D_{i}(t+1)$ are, respectively, the probability and motivation or \textit{drive} of agent $i$ to cooperate at round $t+1$. When the parameter $\beta \geq 0$ is large, agent $i$ tends to cooperate or defect, if the motivation is positive or negative, respectively; if instead the motivation or $\beta$ are zero, the agent acts randomly. Intermediate values of $\beta$ interpolate between these two extremes---rational optimization and random behavior. The term $\Delta U_{i}(t)$ in Eq.~\eqref{eq:d} is the difference in utilities resulting from the choice of either cooperate or defect. If $\Delta U_{i}(t) > 0$, agent $i$'s motivation to cooperate increases; if $\Delta U_{i}(t) < 0$, the motivation decreases, while if $\Delta U_{i}(t) = 0$ it stays the same. Finally, the parameter $\alpha$ describes memory loss: if $\alpha = 1$, the agent remembers only the previous round $t$, while if $\alpha = 0$, the agent has cumulative information of the full history of play. The case of $0 < \alpha < 1$ amounts to an exponential discount of utility over time.

While the EWA model assumes that agents' motivation to cooperate $\Delta 
U_i(t)$ is specified exclusively by individual considerations, e.g., material 
payoffs, in this work we extend the EWA formalism to incorporate normative 
considerations, as described in the next 
section.

\

\tocless\section{\label{s:ewan}Balancing individual and normative considerations}
Here we discuss how we extend the EWA model (see Sec.~\ref{s:learning})
to make agents balance between individual and normative considerations in their 
decision-making. The individual component is common to previous EWA models and implements the assumption of selfishness in Table~\ref{t:assumptions}, while the normative component is introduced in this work and implements the second block of assumptions in Table~\ref{t:assumptions}.
We combine both considerations by defining 
agents' motivation to cooperate $\Delta U_i(t)$ as a weighted sum of the 
individual and normative components. This combination is also consistent with experimental 
observations suggesting that a common area in the brain correlates with the 
computation of both monetary and social rewards~\cite{lin2011social} (see also 
Ref.~\cite{Declerck-Neuroeconomics-2015}). The idea that norms can be conceived 
as part of the utility function that individuals maximize is receiving growing 
attention and empirical 
support~\cite{JEEA:JEEA12006,Vilone-PRE-2014,Declerck-Neuroeconomics-2015,gintis2003hitchhiker,gavrilets2017collective,Bowles-book-2016}.
 In these works, however, social norms typically have an exogenously 
specified impact on individuals' behavior; while we assume that this impact is 
endogenously updated on the basis of how salient, namely relevant, the specific 
norm is perceived to be. 

\

\tocless\subsection{\label{s:decision}Decision rule to cooperate}

Decades of theoretical and experimental work are nowadays putting on solid grounds that when deciding whether to cooperate, humans do not always make choices that maximize their personal payoffs, but they also care about behaving in line with others in their group. Social norms accurately provide information about how members within a certain group will behave and more importantly about how they are prescribed to behave~\cite{fehrFischbacher2004,bicchieri2006,ensmingerHenrich2014}. 

Consistent with this evidence and in analogy with Ref.~\cite{andrighettoEtAl2013}, we develop a model in which the decision makers' utility $\Delta U_{i}(t)$ is based on the material rewards they obtain and on the degree to which their actions comply with social norms. Thus,

\begin{equation}
\label{eq:utility}
\Delta U_{i}(t) = \Delta I_{i}(t) + h \Delta N_{i}(t),
\end{equation}
where the individual drive $\Delta I_{i}(t)$  models the motivation to maximize personal material payoffs and the normative drive $\Delta N_{i}(t)$  models the motivation to comply with social norms (see Sec.~\ref{s:dN} below for more details). The parameter $h$ weights the relative influence of selfish and norm-based motivations on cooperative decision-making: if $h = 0$, agents do not care about normative information, while if $h$ is very large, agents' behavior is dominated by what the norm dictates. It also defines a relative time-scale between selfish and pro-social reasoning, related to reflection and intuition~\cite{Rand-Nature-2012}.

In this way we have incorporated the ability to balance between normative and selfish considerations into the EWA model, keeping the standard EWA formalism almost intact. The impact of both types of considerations on individual decisions has been scarcely explored (but see Refs.~\cite{Bowles-book-2016,Vilone-PRE-2014,Declerck-Neuroeconomics-2015,JEEA:JEEA12006,Rand-Nature-2012}).

\

\tocless\subsection{\label{s:dI}Individual component}
Here we describe the individual component of the model, which implements the assumption of  selfishness in Table~\ref{t:assumptions}. We also describe the specific case of the Prisoner's Dilemma (PD) game because this is the game implemented in the experiment we analyze in Sec.~\ref{s:experiments}. Clearly, other types of games can also be implemented by defining the payoffs accordingly.

We are interested in the situation where agents interact pairwise by playing a given two-player game with each neighbor in a social network~\cite{garcia-lazaroEtAl2012,grujicEtAl2014}. In this case, we can write the individual motivation for agent $i$ to cooperate in her interaction with her neighbor $j$ at round $t$ as ${\Delta I_{ij}(t) = s_j(t)\Delta I_{C} + [1-s_j(t)]\Delta I_{D}}$. Here $s_j(t)$ refers to the strategy played by agent $j$ at round $t$, i.e., whether she cooperated $s_j(t)=1$ or defected $s_j(t)=0$, while $\Delta I_{C}=R-T$ and $\Delta I_{D}=S-P$, where $R$ is $i$'s reward's payoff when both agents cooperate, $P$ is $i$'s punishment's payoff when both agents defect, $S$ is $i$'s sucker's payoff when $i$ cooperates and $j$ defects, and $T$ is $i$'s temptation's payoff when $i$ defects and $j$ cooperates.

The total payoff received by an agent $i$ interacting with $K$ neighbors is given by the average over the payoffs obtained on each of the $K$ pairwise games the agent is involved in. So, the individual drive of agent $i$ to cooperate at round $t$ is 
\begin{equation}\label{e:Idrive}
\Delta I_i(t) = (\Delta I_C-\Delta I_D)\frac{1}{K}n_i(t) +\Delta I_D,
\end{equation}
where  $n_i =\sum_{j\in\partial i} s_j$ refers to the number of $i$'s peers who cooperate, and $\partial i$ stands for the set of neighbors of $i$ in the social network.

If the payoffs satisfy $T > R > P > S$, we have the PD game; furthermore $2 R > 
T+S$ for iterated PD games. The structure of social dilemma is the following: 
although the best individual choice for both is to defect, mutual cooperation 
yields a better payoff than mutual defection ($R > P$). The experiments that we 
analyze here correspond to a \textit{weak} PD game where, in Experimental 
Currency Units (ECUs), $R= 7$ ECUs, $T=10$ ECUs, and $P=S=0$ ECUs; so, $\Delta 
I_C = -3$ ECUs and $\Delta I_D = 0$ ECUs~\cite{garcia-lazaroEtAl2012}.

\

\tocless\subsection{\label{s:dN}Normative component}
Here we describe the normative component of the model which implements the second block of assumptions in Table~\ref{t:assumptions}. We also define the entire utility function [see Eq.~\eqref{eq:utility}] for the specific case of the PD game.

The impact that social norms have on agent's decisions is a function of how salient~\cite{andrighettoEtAl2013,cialdiniGoldstein2004}, i.e., relevant, the norm is perceived by agent $i$ at round $t$ within the social group. The higher the salience of the social norm, the stronger its impact on the motivation to comply with it. The norm salience is determined by two independent factors, weighted with parameters $w_C,\, w_O > 0$, and their interaction, weighted with parameter $w_I > 0$, i.e., (cf.~\cite{andrighettoEtAl2013})

\begin{equation}\label{e:Ndrive}
\Delta N_i(t) = w_C [2s_i(t)-1] + w_O\frac{n_i(t)}{K} + w_I s_i(t)\frac{n_i(t)}{K}.
\end{equation}
According to the first term, the salience of a norm is determined by the behavior at round $t$, namely the agent's choice to comply with or violate the norm. If agent $i$ complies with the norm, she will perceive it as more salient than if she violates it. This is justified by the fact that humans have a strong need to enhance their self-concepts by behaving consistently with their own beliefs and self-ascribed habits so that they can avoid ethical dissonance (self-consistency Refs.~\cite{festinger1957,abelsonBernstein1963,ayalGino2011}). 

Consistent with theoretical and empirical findings on conditional cooperation~\cite{traylsenEtAl2010,fischbacherEtAl2001}, the second term containing $w_O>0$ assumes that the salience of the norm is also affected by the share of peers that complied with it. The more peers comply with the norm, the more salient the norm becomes, and vice-versa. The third term containing $w_I>0$ disappears if the agent did not cooperate at round $t$ (i.e., if $s_i(t)=0$); while not present in Ref.~\cite{andrighettoEtAl2013}, this last term is introduced in this work to account for recent experimental observations that support the MCC rule. , which assumes that in taking decisions individuals are responsive to the behavior of others, but only if they have cooperated themselves~\cite{grujicEtAl2010}. It can be noticed that the second term relaxes the assumption behind the MCC rule by positing that when deciding whether to cooperate  individuals are always sensitive to what others do and not just after having cooperated themselves (social influence). This relaxation is in line to recent findings reported in Ref.~\cite{horitaEtAl2017}.

Comparing Eqs.~\eqref{e:Idrive} and \eqref{e:Ndrive} we can see that the information required about neighbors' action to estimate norm salience, i.e., $n_i(t)$, could in principle be inferred from the information on material payoffs. So, it is not unreasonable to expect that agents can indirectly infer normative information from material payoffs only, as suggested by Horita et al.~\cite{horitaEtAl2017}.

Now, introducing Eqs.~\eqref{e:Idrive}~and~\eqref{e:Ndrive} into Eq.~\eqref{eq:utility}, we get
\begin{equation}\label{e:dUsn}
{\Delta U_i(s_i, n_i) = (a s_i + b) n_i + 2h s_i -h},
\end{equation}
here 
\begin{eqnarray}
a = h w_I/K, \hspace{0.5cm}\hspace{0.5cm} b = (h w_O+\Delta I_{C})/K,
\end{eqnarray} 
are effective parameters introduced to simplify the notation. We have dropped the index $t$ to include explicitly the dependence of $\Delta U_i$ on the number $n_i$ of agent $i$'s peers who cooperated at round $t$.  Furthermore, we have done $w_C=1$, as it can be absorbed in the parameter $h$, and $\Delta I_D = 0$ as we will focus our analysis on the weak PD game studied in Ref.~\cite{garcia-lazaroEtAl2012}.

\ 

\tocless\section{\label{s:network}Slow adaptation and absence of network reciprocity}
Here we describe how to implement the third block of assumptions in Table~\ref{t:assumptions} and how to obtain an effective four-parameter model of a single representative agent. Our interest in such an effective model is that it allows for the complete analytical characterization of its long term dynamics, indicating the existence of critical phenomena, while still quantitatively reproducing major features of large-scale experiments with human groups. 

As already discussed in Sec.~\ref{s:assumptions}, the large-scale experiments analyzed here are consistent with the assumption that adaptation is slow in comparison to the rate of change of individual strategic choices. Fluctuations around agents' average behavior induced by their stochastic nature can then be neglected (see Appendix~\ref{s:1}). This so-called adiabatic~\cite{Galla-PNAS-2013,Sato-PNAS} approximation allows us to replace the stochastic variable $s_i$ encoding the actual strategy chosen by each agent $i$ for its mean value $x_i$, which is a deterministic quantity. 

To see this, notice that introducing Eq.~\eqref{eq:d} into Eq.~\eqref{eq:1} the system dynamics can be fully specified in terms of the cooperation probability as

\begin{equation}\label{e:x-update}
x_i(t+1)=\frac{x_i(t)^{1-\alpha}}{x_i(t)^{1-\alpha}+\left[1-x_i(t)\right]^{1-\alpha}e^{-\beta \Delta U_i(t)}}.
\end{equation}
Replacing the stochastic term $\Delta U_i(t)$, which depends on the actual actions $s_i$ and $s_{\partial i}$ of agent $i$ and her neighbors $\partial i$ (see Eq.~\eqref{e:dUsn}), by its average value $\overline{\Delta U_i}(t)$, which is obtained by changing each action $s_j$ by its corresponding average $x_j$, we get the deterministic equation
\begin{equation}\label{e:x-update-det-nonMF}
x_i(t+1)=\frac{x_i(t)^{1-\alpha}}{x_i(t)^{1-\alpha}+\left[1-x_i(t)\right]^{1-\alpha}e^{-\beta \overline{\Delta U_i}(t)}}.
\end{equation}

More precisely
\begin{equation}\label{e:dut}
\begin{split}
\overline{\Delta U_i}(t) =& \sum_{s_i, s_{ \partial i}}\Delta U(s_i, n_i)\, p_i(s_i, t)\prod_{j\in \partial i}p_j(s_j,t)\\
=& a\, x_i(t)\sum_{j\in \partial i} x_j(t) + b\,\sum_{j\in \partial i}x_j(t)+ 2 h x_i(t) -h,
\end{split}
\end{equation}
where $n_i=\sum_{j\in\partial i} s_j$, the term $s_{\partial i}$ denotes the set of strategies of $i$'s peers, $p_j(s,t)=[x_j(t)]^s[1-x_j(t)]^{1-s}$ is the probability that agent $j$ plays strategy $s$ and we have used the expression in Eq.~\eqref{e:dUsn}.

On the other hand, the interaction structure of a human group does not appear to significantly influence its cooperative behavior~\cite{Sanchez-2015,garcia-lazaroEtAl2012,gutierrez-roigEtAl2014,traylsenEtAl2010,grujicEtAl2014}, this is usually referred to as `absence of network reciprocity'---network reciprocity is the influence of network structure on cooperative behavior~\cite{nowakMay1992}. Meaning that correlations between different agents can then be assumed to be weak. This leads to a mean field approximation~\cite{Mora-StatPhys-2011}, where $\sum_{j\in \partial i} x_j \approx x K$. Here $x$ is the global mean value of $x_i$  calculated over all agents $i$, and $K$ is the average number of neighbors of a generic agent $i$.  This approximation allows us to describe the system in terms of a single representative agent that captures the typical behavior of a generic agent $i$. In this way, we obtain a deterministic learning dynamics of a single representative agent given by the equation (see Appendix~\ref{s:1} and the the first two sections in the supporting information of~\citep{Galla-PNAS-2013} for further details)
\begin{equation}\label{eq:dynamic}
x(t+1)= \frac{x(t)^{1-\alpha}}{x(t)^{1-\alpha}+\left[1-x(t)\right]^{1-\alpha}e^{-\beta\overline{\Delta U}[x(t)]}},
\end{equation} 
where $x$ is the probability for the representative agent to cooperate, and
\begin{equation}\label{eq:mfUx}
\overline{\Delta U}[x]= a K\, x^2 + (b K + 2 h)\,x -h,
\end{equation}
is obtained by replacing in Eq.~\eqref{eq:utility} both $s_i$ and $n_i/K$ with the average value $x$. Equation~\eqref{eq:dynamic} describes the relevant aspects of the dynamics of the global cooperation level and can reproduce the values observed in Ref.~\cite{garcia-lazaroEtAl2012} with accuracy comparable to more complex models~\cite{ciminiSanchez2014,villatoroEtAl2014} (see Sec.~\ref{s:experiments}).

\ 

\tocless\section{\label{s:dynamic}Dynamical regimes and phase diagram}
Here we determine the phase diagram characterizing the long-term dynamics of 
the effective single-agent model described in detail in the previous sections; 
this diagram shows three regimes---mono-stability, bi-stability, and 
non-equilibrium---as well as a line of critical points.

To study the long-term dynamics of the model defined in 
Eq.~\eqref{eq:dynamic} we look for fixed points, i.e., points $x$ that satisfy 
$x(t+1)=x(t)=x$. The points at the boundary, i.e., $x=0$ and $x=1$, are 
fixed points of the mean field dynamics described by Eq.~\eqref{eq:dynamic}, 
but they are unstable since $\alpha,\,\beta >0$. Only fixed points $x^\ast$ 
satisfying $0<x^\ast<1$ can be stable. The condition that these points satisfy 
can be derived from Eq.~\eqref{eq:dynamic} by doing $x(t+1)=x(t)=x$, which 
yields
\begin{equation}\label{e:f}
x=f(x)\hspace{0.3cm}\textrm{with}\hspace{0.3cm} f(x) =\frac{1}{2}+\frac{1}{2}\tanh\left[A(x-x_0)^2+y_0\right],
\end{equation}
where
\begin{eqnarray}
A &=& \frac{a K}{2 \gamma} ,\label{e:A}\\
x_0 &=& -\frac{b K + 2 h}{2\, a K} ,\label{e:x0}\\
y_0 &=& -\frac{(b K + 2 h)^2}{8\, a K \gamma} -\frac{{h}}{2 \gamma} ,\label{e:y0}
\end{eqnarray}
and $\gamma = \alpha/\beta$. In Eq.~\eqref{e:f} we have not made explicit 
the dependence of the function $f$ on the effective parameters $A$, $x_0$, and 
$y_0$ to reduce clutter in the notation.

If the MCC assumption is dropped, i.e., $w_I=0$ so $a = 0$, Eq.~\eqref{e:f} 
becomes equivalent to the equation that determines the equilibrium 
magnetization, given by $m = 2 x -1$, of the Curie-Weiss 
model~\cite{Sethna-book-2006}. Indeed, when $a = 0$ Eq.~\eqref{e:f} can be 
written as ${m = \tanh\left[\beta (J_{\rm eff}m + H_{\rm eff})\right]}$, where 
${J_{\rm eff} = (h w_O + \Delta I_C + 2h)/4\alpha}$ would correspond to an 
effective ferromagnetic interaction (when $J_{\rm eff}>0$) and ${H_{\rm eff} = 
(h w_O + \Delta I_C)/4\alpha}$ would correspond to an effective external field. 
As it is well known, the Curie-Weiss model can display two phases, paramagnetic 
and ferromagnetic, which are the magnetic analogous of the regimes of 
mono-stability and bi-stability of our model of human cooperative dynamics.

The MCC assumption ($w_I >0$) introduces an additional non-linearity, whose 
magnetic analogous is an additive term of order $m^2$ in the argument of the 
hyperbolic tangent. Such term comes from the interaction between the 
agent's own cooperative behavior and that of her neighbors. Such additional 
non-linearity renders the phase diagram of the model more complex and gives 
rise to a new non-equilibrium phase, where the cooperative dynamics never 
settles.

Indeed, as described in detail in Appendix~\ref{s:phase} and shown in 
Fig.~\ref{fig:criticality}, depending on the values of the parameters $A,\, 
x_0,\, y_0$ there can be zero, one, or two stable fixed points, corresponding 
to a non-equilibrium, mono-stable, or bi-stable long-term dynamics, 
respectively. This provides an analytical characterization of the system that 
helps to obtain insights about systems as complex as human groups that are 
typically difficult to obtain. In particular, this analytical 
characterization allows us to infer model parameters from experimental data and 
identify evidence that the human groups playing in the experiments of 
Ref.~\cite{garcia-lazaroEtAl2012} are near criticality (see 
Sec.~\ref{s:experiments} and Appendix~\ref{s:inference}). The strategy we 
adopt here for the estimation of model parameters from experiments uses 
information about the dynamical regimes identified. 

\begin{figure*}
	\centering
	\includegraphics[width=\textwidth]{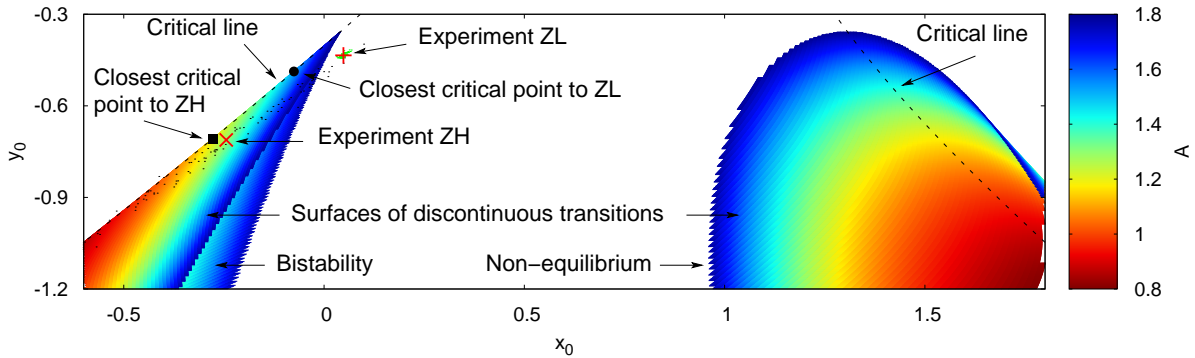}
	\caption{\label{fig:criticality}{Dynamical regimes of human cooperation}. Color map representation of the surfaces of discontinuous transitions defined by a function $A_d(x_0,y_0)$ that returns the transition vale of $A$ for each value of the effective parameters $x_0$ and $y_0$, defined in Eqs~\eqref{e:x0} and \eqref{e:y0}, respectively. Each color encodes a level curve $A_d(x_0,y_0)=A$, which partitions the $x_{0}$-$y_{0}$ plane into three regions corresponding to different long-term dynamical regimes: Inside the sharp triangular-like curve (left) the system is bistable; inside the parabolic-like region (right) the system never reaches equilibrium. Outside these two regions the system is monostable (Appendix~\ref{s:phase}). The cloud of black and green dots represent a projection on the $x_0$-$y_0$ plane of the posterior population of parameters $\alpha$, $A$, $x_0$, and $y_0$ inferred here from the experiments performed in Zaragoza~\cite{garcia-lazaroEtAl2012} on an heterogeneous network (ZH) and on a lattice (ZL), respectively. We show the parameters estimated (ZH: diagonal cross; ZL: vertical cross) and quantify its relative distance (ZH: 3\%; ZL: 11\%) to the closest point (ZH: square; ZL: circle) on the critical lines (dashed) (Appendix~\ref{s:inference}).}
\end{figure*}

The regions of the phase diagram corresponding to the different dynamical regimes are separated by surfaces of discontinuous transitions that terminate on a line of critical points (see Appendix~\ref{s:phase}). At these critical points correlations are known to become long-range~\cite{Mora-StatPhys-2011} and systems have been shown to display a multitude of significant features, like large repertoire of dynamical responses, optimal transmission and storage of information, and extreme sensitivity to external perturbations~\cite{hidalgoEtAl2014,Mora-StatPhys-2011,gelblumEtAl2015,Tkacik,Chate-Physics-2014,bialekEtAl2014,attanasiEtAl2014,Krotov2014,Nykter12022008,Mora2009,Beggs2008}. 

Several mechanisms have been put forward in an attempt to explain how criticality could emerge in living systems~\cite{Chate-Physics-2014}. A novel perspective posits that criticality is the evolutionary stable outcome of a group of individuals equipped with mechanisms aimed at representing each other with fidelity, wherein the best possible trade-off between accuracy and flexibility is achieved~\cite{hidalgoEtAl2014}. We here show evidence that mechanisms balancing between individuality and social conformity can underlie human cooperation.

Criticality is usually associated to the divergence of a properly defined susceptibility that quantifies the range of the correlations in the system and its response to external perturbations~\cite{Mora-StatPhys-2011,Sethna-book-2006}. Here it can be defined in terms of the change in the global cooperation $x$ when a certain model parameter $\theta$ varies, e.g., $\theta =h$. Out of the non-equilibrium region it is given by (see Appendix~\ref{s:cExp})
\begin{equation}
\frac{\partial x}{\partial\theta} \propto \frac{1}{A_c^\ast-A},
\end{equation}
which clearly diverges when $A$ approaches a critical point, described here by $A_c^\ast$; notice that $A$ varies with the original parameters of the model, since it is defined in terms with them [see Eq.~\eqref{e:A}].

In the concluding section, we discuss the implications that this characteristic may have for the adaptiveness of human groups.

\ 

\tocless\section{\label{s:experiments}Analysis of large-scale experiments of humans playing a Prisoner's Dilemma}
Here we use experimental data from Ref.~\cite{garcia-lazaroEtAl2012} to 
determine the parameters of the effective single-agent model described in 
Secs.~\ref{s:assumptions}-\ref{s:network} and locate the human group playing in 
the experiment into the phase diagram obtained in Sec.~\ref{s:dynamic} (see 
Fig.~\ref{fig:criticality}).

\

\tocless\subsection{Brief review of experiments analyzed}
To estimate where human groups may locate in the phase diagram of 
Fig.~\ref{fig:criticality}, we extracted the model parameters from two recent 
large-scale experiments in which more than 600 human participants play 
simultaneously a Prisoner's Dilemma game in two different network 
environments~\cite{garcia-lazaroEtAl2012}. These experiments are aimed at 
testing the relative effect of homogeneous or heterogeneous networks 
environments on cooperative behavior (for details see 
Appendix~\ref{s:inference}). We build on these experiments because we expect 
them to offer more robust statistics than similar, but smaller experiments. 

In Ref.~\cite{garcia-lazaroEtAl2012} one of the two experiments was conducted 
on a square lattice and the other on a heterogeneous network. However, their 
finding that network structure does not significantly affect behavior (i.e., 
the absence of network reciprocity) suggests that even though our mean field 
model neglects network structure, it can still provide a good description of 
the experiments, as shown below. In these experiments, human subjects played a 
$2 \times 2$ multi-player PD game with each of their $K$ neighbors for $52$ 
rounds. Players could take only one action---either to cooperate (C) or defect 
(D)---the action being the same against all opponents. The experiment was 
simultaneously carried out on two different virtual networks: the first network 
consisted in a $25 \times 25$ lattice with a fixed number of 4 neighbors and 
periodic boundary conditions (625 subjects); the second network was a 
heterogeneous network with a fat-tailed degree distribution (604 subjects), 
where the number of neighbors varied between 2 and 16. 

Subjects played a repeated (weak) Prisoner's Dilemma game with all their neighbors for an initially undetermined number of rounds. Payoffs  were set to be 7 Experimental Currency Units (ECUs) for mutual cooperation, 10 ECUs for a defector facing a cooperator, and 0 ECUs for any player facing a defector. 

Participants received information about the actions and normalized payoffs of their neighbors in the previous round. Without knowledge of the duration of the game, participants had to make only one decision for all neighbors. Therefore, the situation becomes similar to a repeated public goods game. In public goods experiments, participants usually start highly cooperative, but in the absence of cooperation-enhancing mechanisms, such as punishment or reputation, their cooperation levels decreases over time. Information about the behavior of others allows participants to create expectations about how others will behave, namely about the social norms ruling the group.

\begin{figure*}
\centering\includegraphics[width=0.9\textwidth]{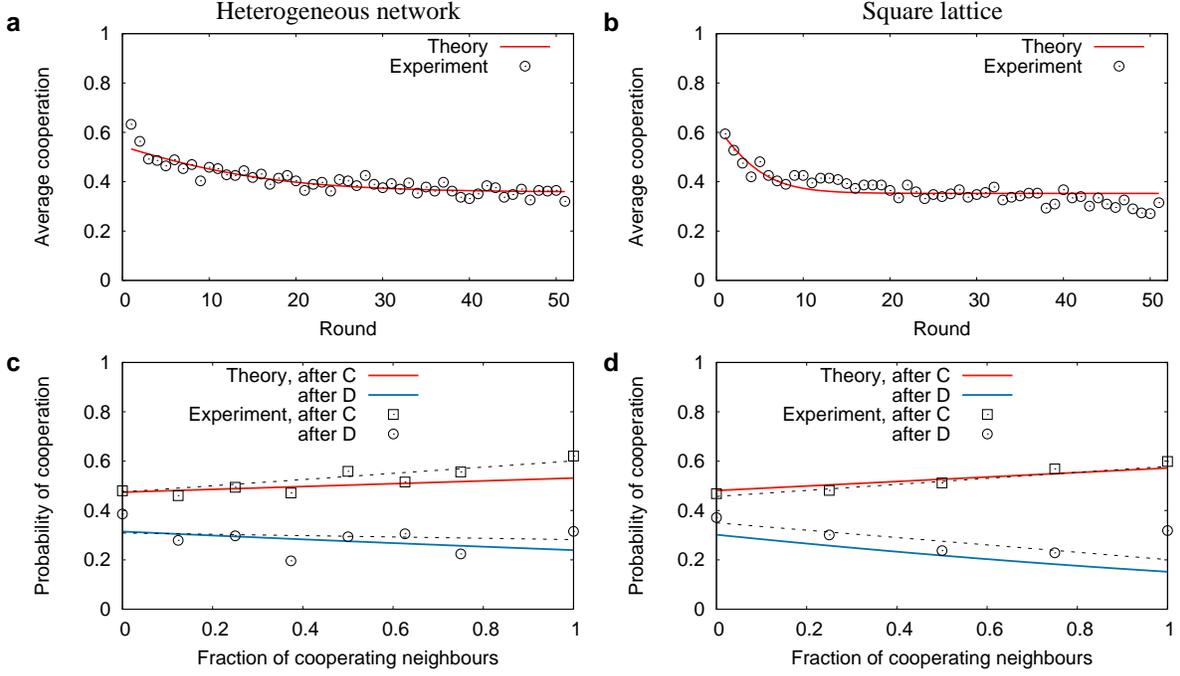}
\caption{\label{fig:results}{Balance of selfishness and conformity to social norms explains human behavior in large-scale Prisoner's Dilemma games} {(a, b)}. Comparison of the dynamics of the global cooperation level observed in the laboratory experiment conducted in Zaragoza~\cite{garcia-lazaroEtAl2012} (circles) on an heterogeneous network (a) and on a square lattice (b) with that predicted by Eq.~\eqref{eq:dynamic} (line) with the corresponding parameters inferred from the same experiments (see caption of Fig.~\ref{fig:criticality}). The model and experimental dynamics are in  agreement even in the transient regime. {(c, d)} Probability for a representative agent to cooperate in a generic round based on whether she cooperated (C, squares) or defected (D, circles) and on the number of neighbors who cooperated in the previous round obtained in the same experiments~\cite{garcia-lazaroEtAl2012} on an heterogeneous network (c) and on a lattice (d). The experimental data is compared with the values predicted by Eq.~\eqref{eq:prob} (lines) with the corresponding parameters inferred from the same experiments (see caption of Fig.~\ref{fig:criticality}). We can see that the assumption of linearity is valid and that our model agrees with the experimental values to a large extent. We include the linear fits  (dashed straight lines) directly obtained from experimental data~\cite{garcia-lazaroEtAl2012} for comparison.}
\end{figure*}

We focus here on two features observed in these experiments that can be 
reproduced by our model (Fig.~\ref{fig:results}). The first feature is the 
dynamics of the global cooperation level, which decays from an initial value of 
about 60\% to a relatively constant value of about 35\%, both on the 
heterogeneous network and on the square lattice [Fig.~\ref{fig:results}(a) and 
(b)]. The second feature is the probability ($P$) for a generic individual to 
Cooperate ($C$) or to Defect ($D$) in a round, conditioned by her previous 
action $s$ and the number $n$ of neighbors who cooperated in the previous 
round~\cite{Sanchez-2015,garcia-lazaroEtAl2012,gutierrez-roigEtAl2014,traylsenEtAl2010,grujicEtAl2014}. We denote this probability by $P \left( C | s, n \right)$. Ref.~\cite{garcia-lazaroEtAl2012}, for instance, reports a nearly linear dependence of $P \left( C | s, n \right)$ from $n$, for both values of $s$ [see Fig.~\ref{fig:results}(c) and (d)].

\

\tocless\subsection{Inference of model parameters}
To fit our model to the experimental data, we notice that the left hand side of Eq.~\eqref{eq:1} for the representative agent can be interpreted as ${x(t+1) = P \left(C, t + 1 | s, n, x, t \right)}$, namely the probability that the agent cooperates at round $t+1$ given that at round $t$ the following three conditions are satisfied: (i) she played strategy $s$, (ii) $n$ of her neighbors cooperated, and (iii) $x(t)=x$. To eliminate the explicit dependence on the history of the system, i.e., on $x$ and $t$, we first assume that the system of interacting humans observed in the laboratory has reached a stationary state, so that it can be accurately described by the long-term mean field dynamics [Eq.~\eqref{eq:dynamic}]. We further assume that the system is essentially mono-stable and, accordingly, the model dynamics is dominated by a single fixed point. These assumptions considerably simplify the analysis and, as it turns out, are self-consistent with the results obtained (see Appendix~\ref{s:inference} for a detailed treatment). 

Although the stationary state of a generic system can depend on its dynamical history, under the above assumptions this is not the case. Thus, the right hand side of Eq.~\eqref{eq:1} evaluated at the fixed point, i.e.,
\begin{equation}
\label{eq:prob}
P \left( C | s,n \right) = \frac{1}{1 + y_{1}^{1 - \alpha} e^{-\beta\Delta U \left( s,n \right)}},
\end{equation}
should coincide with the experimental results, where $y_{1} = \left( 1 - x_{1} \right) / x_{1}$, and $x_{1}$ is the global cooperation level at the dominant stable fixed point.
The term $\Delta U(s,n)$ is the utility function of the representative agent in the mean field approximation, which is obtained by simply dropping the agent index $i$ in Eq.~\eqref{e:dUsn}. Similarly, ${y_{1} = \left( 1 - x_{1} \right) / x_{1}}$, with $x_{1}$ being the only stable fixed point of Eq.~\eqref{eq:dynamic}. This result indicates that when the system is deterministic and monostable its long-term dynamics is independent of its history. When the system is bistable and we neglect fluctuations altogether, the probability $P ( C | s,n )$ is given instead by a convex combination of terms like the one on the right hand side of Eq.~\eqref{eq:prob}, one for each fixed point. More details can be found in Appendix~\ref{s:mcc}.

To compare Eq.~\eqref{eq:prob} with the nearly linear behavior [Figs.~\ref{fig:results}(c) and~\ref{fig:results}(d)] observed in~\cite{Sanchez-2015,garcia-lazaroEtAl2012} (but see Ref.~\cite{gutierrez-roigEtAl2014}), we do a first order approximation in $\beta$ to obtain
\begin{equation}\label{eq:linear_prob}
{P \left( C | s,n \right) = m_{s} n/K + r_{s}}, 
\end{equation}
with 
\begin{eqnarray}
m_s &=& \beta K J(\alpha)(a s + b),\label{e:mcc-1fpA_DEF}\\
r_s &=& I(\alpha)+\beta J(\alpha)\left[h(2 s- 1) \right]\label{e:mcc-1fpB_DEF};
\end{eqnarray}
this approximation is consistent with the results below (see Fig.~\ref{fig:results}). 

The slopes $m_s$ and intercepts $r_s$ in Eq.~\eqref{eq:linear_prob} are better described in terms of the mean intercept $r$ and the `gap' $G$ between intercepts of the near linear trends that describe the MCC rule~\cite{garcia-lazaroEtAl2012}, i.e.,
\begin{eqnarray}
r &=& \frac{1}{2}(r_C+r_D) = I(\alpha),\label{e:r_DEF}\\
G &=& r_C - r_D = 2 \beta h w_C J(\alpha),\label{e:G_DEF}
\end{eqnarray}
where, for convenience, we have done $m_0 \equiv m_D$, $r_1 \equiv r_C$, etc.; here
\begin{eqnarray}
I(\alpha) \equiv \frac{1}{1+y_1^{1-\alpha}},\\
J(\alpha) \equiv  \frac{y_1^{1-\alpha}}{\left(1+y_1^{1-\alpha}\right)^2}.
\end{eqnarray}

In experiments we have $m_D\neq 0$, which implies $b\neq 0$ as $m_D\propto b$, so
\begin{eqnarray}
m_C-m_D &=& \beta\, a K J(\alpha),\label{e:dm_DEF}\\
\frac{m_C}{m_D} &=& \frac{\beta\, a+\beta\, b}{\beta\, b}.\label{e:mCmD_DEF}
\end{eqnarray}

We are now in a better position to discuss the role played by the parameters $w_C,\, w_O,\, w_I$ encoding the normative assumptions. First, notice that we have re-introduced parameter $w_C = 1$ in Eq.~\eqref{e:G_DEF} to make explicit that if the assumption of self-consistency (see Table~\ref{t:assumptions}) were dropped, i.e., if we did $w_C = 0$, the gap $G$ would vanish, contradicting experimental observations (see Fig.~\ref{fig:results}). Analogously, we can see from Eq.~\eqref{e:dm_DEF} that if the MCC assumption (see Table~\ref{t:assumptions}) were dropped, i.e., if we did $w_I = 0$ which implies $a=0$, the two slopes $m_C = m_D$ are equal, contradicting experimental observations (see Fig.~\ref{fig:results}). In this sense, parameters $w_I$ and $w_C$ play not only a quantitative but also a qualitative role. In contrast, the role of parameter $w_O$ is more quantitative than qualitative. Indeed, Eq.~\eqref{e:mcc-1fpA_DEF} implies that ${m_D = \beta K J(\alpha) b}$. If $w_O = 0$ we have $b = \Delta I_C /K$, which is fixed by the experimental conditions, and $m_D = \beta J(\alpha ) \Delta I_C $ would be less than zero for the PD game, since $\Delta I_C <0$; this is consistent with experimental observations. However, with $w_O = 0$ the accuracy of the fit was rather poor and so we did not include it in our analysis. However, future analysis should study in further detail the relevance of this assumption.

As described in detail in Appendix~\ref{s:mcc_mono}, there is a direct relationship between the parameters of the model and the experimental quantities defined above [see Eqs.~\eqref{e:alphaexp}-\eqref{e:bexp0}].  So, the values of $m_s$ and $r_s$, extracted from experimental data~\cite{garcia-lazaroEtAl2012} (see Table~\ref{t:data}), constraint the values of the model parameters. There is a further constraint: the dynamics of the global cooperation level should be consistent with experimental results [Figs.~\ref{fig:results}(a) and~\ref{fig:results}(b)].

A population of parameters satisfying the resulting set of constraints was obtained via Bayesian inference by using the package \texttt{pomp}~\cite{pomp} and is illustrated in Fig.~\ref{fig:criticality}. Although the 2D projection of the phase diagram in Fig.~\ref{fig:criticality} may suggest otherwise, they all lie on the region of mono-stability. The technical details and the data obtained are provided in Appendix~\ref{s:inference}. 

\

\tocless\subsection{Results}
The parameters corresponding to the two experiments (see Table~\ref{t:diagramZH}), inferred by the method described above (see also Appendix~\ref{s:inference}), are at a relative distance of 3\% and 11\% to the critical line (see Fig.~\ref{fig:criticality} and Appendix~\ref{s:inference}). 

Figures~\ref{fig:results}(a) and~\ref{fig:results}(b) compare the levels of global cooperation observed in the laboratory experiment~\cite{garcia-lazaroEtAl2012} (circles) with the ones predicted by Eq.~\eqref{e:x-update} (line), informed with values extracted from Ref.~\cite{garcia-lazaroEtAl2012}. 
Results from both the heterogeneous [Fig.~\ref{fig:results}(a)] and homogeneous [Fig.~\ref{fig:results}(b)] networks are presented. Both figures show a decay in cooperation over the 52 rounds from the initial value of 60\% to around 35\% in both treatments. Results show a close agreement of the model dynamics with the laboratory experiments. Likewise in Ref.~\cite{garcia-lazaroEtAl2012}, the network topology does not have any appreciable influence in the evolution of the level of cooperation.

Figures~\ref{fig:results}(c) and~\ref{fig:results}(d) show the probability for a representative agent to cooperate in a generic round based on whether she cooperated (C, squares) or defected (D, circles) and on the number of neighbors who cooperated in the previous round. Results obtained in both the heterogeneous network [Fig.~\ref{fig:results}(c)] and lattice [Fig.~\ref{fig:results}(d)] are shown. Again, both figures indicate that the probability defined in Eq.~\eqref{eq:prob} is consistent with both the experiments and the linear approximation in Eq.~\eqref{eq:linear_prob}. 

Our model reproduces human cooperative behavior observed in large-scale laboratory experiments more accurately than the MCC behavioral rule, since, as shown in~\cite{grujicEtAl2010,garcia-lazaroEtAl2012}, the latter is not able to reproduce the slow decay of the cooperation level when the agents did not cooperate in the immediate past.

\ 

\tocless\section{\label{s:conclusion}Conclusion}
In this work we presented a statistical physics based model to account for human decision processes behind cooperative behavior. In this model, the decision makers' utility is based both on the material rewards they obtain and on the degree to which their actions comply with social norms.  Results from this analytically tractable model are in agreement with observations from recent large-scale experiments with humans~\cite{garcia-lazaroEtAl2012}.
The model closely reproduces both the global cooperation level and the final distribution of agents according to their probability of cooperation. This provides support to our hypothesis that human cooperation is the outcome of the interaction between instrumental decision-making, aimed to maximize people's economic rewards, and the norm psychology humans are endowed with. In doing so, we have provided experimental evidence of the effect of social norms in promoting cooperative behavior in large groups of humans facing a social dilemma situation.

The cognitively inspired model presented  encapsulates important empirical knowledge on human cooperative behavior: i) humans' social strategic behavior operates with both model-free and model-based reinforcement learning~\cite{camererHo1999,Glimcher-book-2013} that are at the basis of the EWA framework adopted; ii) population structure does not significantly influence the cooperative outcome~\cite{Sanchez-2015,garcia-lazaroEtAl2012,gutierrez-roigEtAl2014,traylsenEtAl2010,grujicEtAl2014} that in the model led to a mean field approximation; 
iii) adaptation is slow when compared with the time scale at which individual actions change, which allows us to neglect in the model stochastic fluctuations and obtain a deterministic dynamics [see Appendix~\ref{s:1} and Eq.~\eqref{eq:dynamic}].

The presented model is parsimonious enough to allow for a detailed characterization of its long-term dynamics. By inferring the model's parameters from experimental data extracted from Ref.~\cite{garcia-lazaroEtAl2012}, we show that the cooperative system is located near criticality.

Recently, evidence has been mounting that living systems, like human brain, insect swarms, gene expression networks, bird flocks, and fish schools~\cite{Mora-StatPhys-2011,Chate-Physics-2014,gelblumEtAl2015,bialekEtAl2014,Tkacik} 
operate near critical points and this might provide them functional advantages. Far from criticality a system can be either too stable, which may favor maladaptive behaviors, or too uncoordinated with its members behaving essentially interdependent of each other. In both extremes the system as a whole is not very responsive to external changes, while around a critical point it is strongly correlated and highly sensitive to changes and its capacity to respond efficiently to varying external conditions can be maximized~\cite{hidalgoEtAl2014}. 

Even though still preliminary, our evidence of signatures of criticality in human cooperative groups is in agreement with recent findings on socio-ecological systems showing that social norms enhance the adaptiveness of cooperative systems to social and environmental variability~\cite{schluterEtAl2016,ostrom2005}. These studies report that during times of institutional and ecological volatility, social norms facilitates the management of common resources like forests, water, fisheries, more than the action of formal institutions. The long-range correlations between pairs of human subjects associated to a critical point could then help explain why norm-based cooperation may enhance the adaptiveness of human groups to external change. Social norms are then crucial mechanisms for both promoting cooperation and enhancing its resilience to external perturbation. 

Clearly, more theoretical and empirical work is needed to reach solid conclusions. For example, machine learning techniques, like the maximum entropy approach in Refs.~\cite{Tkacik,Brunton-PNAS-2016,Mora-PRL-2015,Cavagna-PRE-201}, can be used to carry out a complementary data-driven analysis that does not rely on expert knowledge as the model we presented here. Moreover, experiments that vary some of the relevant parameters of the model, e.g., the payoff matrices, specifically targeted to more directly address our findings need to be performed. 

However, the increasing number of similar evidence~\cite{Chate-Physics-2014,bialekEtAl2014,gelblumEtAl2015,attanasiEtAl2014,Mora-StatPhys-2011,Tkacik,Nykter12022008,Mora2009,Beggs2008} attesting criticality in living systems seems to support the plausibility of our results. Similarly to ants~\cite{gelblumEtAl2015}, human groups appear to reach optimal coordination at a suitable trade-off between individuality and social conformity~\cite{gelblumEtAl2015}, and this makes them to poise at the critical point. 
Social conformity increases the ability of a group to coordinate to reach the desired collective outcome. However, behavioral conformism has also the disadvantage of increasing the stability of undesirable behaviors and of decreasing the ability of the system to react to external information~\cite{gelblumEtAl2015}. Thus the optimal collective performance is achieved when group members are able to balance between social conformism and individuality, so that they are able to achieve a high level of coordination within the group, but also to maintain a robust responsiveness to external perturbations. 

How would humans tune to criticality? An intriguing possibility is that humans implicitly build a model and adjust its parameters accordingly, similar in a sense to what we did here. Model-based inference techniques apparently tend to produce parameter values that are close to a critical point~\cite{mastromatteoMarsili2011}. Model-based learning mechanisms in humans~\cite{Lee-Cell-2016,camererHo1999} could then influence their behavior and drive human groups towards criticality~\cite{hidalgoEtAl2014}. Human subjects hardly possess global structural information about their group, which may explain why the mean field model developed here is accurate enough, and ultimately why no significant impact of population structure on cooperative behavior has been observed~\cite{Sanchez-2015}. An alternative idea~\cite{attanasiEtAl2014,gelblumEtAl2015} posits that biological groups can tune to criticality by growing until a suitable size. If so, it may be difficult to observe signatures of criticality in experimental setups with human groups of fixed size. 

Another interesting question that arises is whether there may be a connection between the signatures of criticality observed here in a group of decision makers and those that have been observed in the brains powering the decision making itself~\cite{Tkacik,Mora-PRL-2015,Beggs2008}. 

\ 

\tocless\section{Acknowledgments}
We thank Carlos Gersherson for facilitating this collaboration through the FuturICT Latin American Node. We also thank Alejandro Perdomo-Ortiz, Marcello Benedetti, Luca Tummolini, and Daniele Vilone for useful discussions.
GA was partially supported by the Knut and Wallenberg Grant ``How do human norms form and change?'' 2016-2021 and by the European Union's Horizon 2020 Project PROTON under Grant agreement no.: 699824.

\

\tocless\section{Author contributions}
All authors contributed to the design of the research, the development of the model, and the writing of the manuscript. JRG did the mean field analysis, the analytical calculations, the Bayesian parameter estimation, and led the research. JRG and GA wrote the first draft of the manuscript. JAM contributed to algorithm development and data analysis.

\


\appendix
\noappendicestocpagenum
\tableofcontents
\section{\label{s:1}Slow adaptation and adiabatic approximation}
A way to justify the approach leading to Eqs.~\eqref{e:x-update-det-nonMF} and \eqref{e:dut} in the main text is assuming that the cooperation probability, or equivalently the drive in Eq.~\eqref{eq:d}, changes slowly during a batch of about $T$ rounds~\cite{Galla-PNAS-2013,galla2009,Realpe-JSTAT-2012}. For small values of $T$, a linear noise correction to the deterministic equation give good results even for a number of players as small as two~\cite{galla2009}. Since we are interested here in games with hundreds of players and we are focusing exclusively on observed experimental features at the aggregate level, namely the global level of cooperation and the MCC rule, we take $T=1$ and neglect the noise altogether. This approach is expected to be better suited for games with a sufficient large number of agents and is not expected to necessarily described the initial transient regime in sufficient detail. As discussed in the main text, this approach can actually describe the major features of the largest experiment to date~\cite{garcia-lazaroEtAl2012} with enough qualitative and quantitative detail.

If $\alpha$ and $\beta$ both vanish, the probability of cooperation remains constant. We will thus assume that $\alpha $ and $\beta$ are small so that changes in the drive during a few rounds are not appreciable. The accumulated changes will then only become noticeable after each batch of $T$ rounds; these can be written as
\begin{equation}\label{e:DitT}
D_i(t+T) = (1-\alpha)^T D_i(t) + \sum_{k=0}^{T-1}(1-\alpha)^{k}\Delta U_i(t+k).
\end{equation}
Here the sum is over the $T$ consecutive rounds that start at round $t$. We can re-write Eq.~\eqref{e:DitT} as
\begin{equation}\label{e:DT}
\frac{D_i(t+T)}{T} = (1-\alpha^\prime) \frac{D_i(t)}{T} + \frac{1}{T}\sum_{k=0}^{T-1}(1-{\alpha^\prime})^{k/T}\Delta U_i(t+k),
\end{equation}
where $\alpha^\prime = 1- (1-\alpha)^T\approx \alpha T$, since $\alpha$ is assumed small.

For large values of the batch size $T$, we can interpret the sum in the right hand side of Eq. \eqref{e:DT} as a weighted time average. The weight is given by a discounting factor $(1-\alpha^\prime)^{k/T}$ which decreases from $1$ to $1-\alpha^\prime$ from the beginning ($k=0$) to end ($k=T-1$) of the batch, respectively. So, if we further assume that also $\alpha^\prime\approx \alpha T$ is small, we can approximate such a sum by the ensemble average in Eq.~\eqref{e:dut} calculated with the corresponding mixed strategies~\cite{Galla-PNAS-2013,galla2009,Realpe-JSTAT-2012}. 

Replacing the last term in Eq.~\eqref{e:DT} with the term defined in Eq.~\eqref{e:dut} and writing everything in terms of a rescaled time $\tau \equiv t/T$ and a rescaled drive $D_i^\prime(\tau) = D_i(\tau\, T)/T$ and utility differences $\Delta U^\prime_i(\tau) = \overline{\Delta U_i}(\tau\, T)/T$, we obtain
\begin{equation}\label{e:Ditau}
{D^\prime_i(\tau+1)} = (1-\alpha^\prime) {D^\prime_i(\tau)} + {\Delta U_i^\prime(\tau)}. 
\end{equation}

Following Eqs.~\eqref{eq:1}~and~\eqref{e:Ditau}, and the definitions $\tau \equiv t/T$ and $x^\prime(\tau)\equiv x(\tau T)$, we can write 
\begin{equation}
x(t+T) = x^\prime(\tau+1)=\frac{1}{1+e^{-\beta D^\prime_i(\tau +1)}}. 
\end{equation}
In terms of a rescaled parameter $\beta^\prime=\beta T$, we obtain an equation analogous to Eq.~\eqref{e:x-update} but for updates on batches of $T$ rounds, i.e.
\begin{equation}\label{e:x-det}
x^\prime_i(\tau+1) = \frac{x^\prime_i(\tau)^{1-\alpha^\prime}}{x^\prime_i(\tau)^{1-\alpha^\prime}+[1-x^\prime_i(\tau)]^{1-\alpha^\prime}e^{-\beta^\prime{\Delta U_i^\prime(\tau)}}},
\end{equation}
where, introducing Eq.~\eqref{e:dUsn} into Eq.~\eqref{e:dut}, for the case of the Weak Prisoner's Dilemma we are interested in we have
\begin{equation}\label{e:dUprime}
{\Delta U_i^\prime(\tau)} = a x_i^\prime (\tau)\sum_{j\in \partial i} x_j^\prime(\tau) + b\sum_{j\in \partial i}x_j^\prime(\tau)+ 2 h x_i^\prime(\tau) -h.
\end{equation}
Eq.~\eqref{e:x-det} is a deterministic update rule obtained by neglecting the fluctuations in the last term in Eq.~\eqref{e:DT}, which is stochastic, and replacing it with the average in Eq.~\eqref{e:dut}. Finally, notice that since we assumed $\alpha^\prime$ is small then $\alpha\approx \alpha^\prime / T$ should be even smaller. Notice also that in this case the ratio $\alpha^\prime /\beta^\prime =  \alpha / \beta$ remains the same. 

If we take $T=1$ then $\tau = t$, so we can replace $x^\prime_i(\tau)\to x_i(t)$ and $\Delta U^\prime (t)\to\overline{\Delta U_i}(t)$ in Eq.~\eqref{e:dUprime}. Similarly, we can replace $\alpha^\prime\to \alpha$ and $\beta^\prime\to\beta$. This yields Eqs.~\eqref{e:x-update-det-nonMF}~and~\eqref{e:dut} above.

\section{\label{s:phase}Calculation of the phase diagram}
Here we show that Eq.~\eqref{eq:dynamic} indeed predicts three regimes with qualitatively different long-term dynamics: mono-stable, bi-stable, and non-equilibrium. 

Graphically, the solutions of Eq.~\eqref{e:f} correspond to the intersections of the graphs of $f$ and the identity function at points $x$ that satisfy $0< x <1$. Their stability is determined by the magnitude of the derivative of $f$, i.e.
\begin{equation}\label{e:dfdx}
f^\prime(x)\equiv \frac{\partial f(x)}{\partial x} = \frac{{A}\, (x-x_0)}{\cosh^2\left[A(x-x_0)^2+y_0\right]},
\end{equation}
evaluated at the corresponding intersection point $x$: If $|f^\prime(x)| < 1$ (respectively, $|f^\prime(x)| > 1$), then the fixed point is stable (respectively, unstable). We have used partial rather than total derivative in Eq.~\eqref{e:dfdx} to stress that $f$ is also a function of $A$, $x_0$ and $y_0$.

\begin{figure}\centering
\includegraphics[width=1\columnwidth]{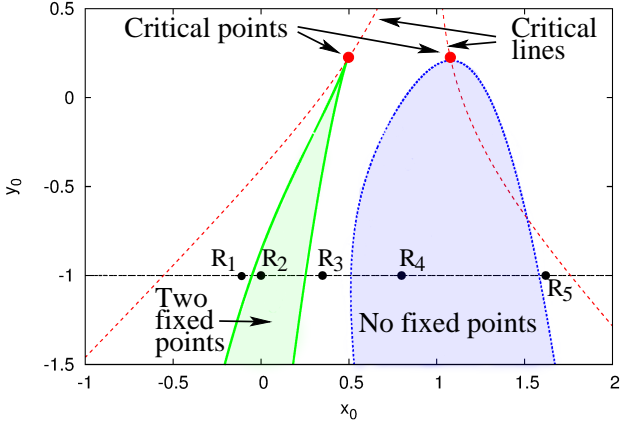}
\caption{A level curve of the surface $A_d(x_0, y_0)$ of discontinuous transitions (see Eqs.~\eqref{e:Atzy0} and~\eqref{e:x0zy0}), i.e., $A_d(x_0, y_0) = A$ (here $A=5$). Inside the green triangular-like region (left) there are two stable fixed points. Inside the blue parabolic-like region (right) there are no stable fixed points. In all the remaining white area there is one stable fixed point. These regions terminate on critical points (red circles). If the value of $A$ changes, these regions shift and so do the corresponding critical points along the critical lines (red dashed lines). The points $R1,\dotsc, R_5$ on the black dashed horizontal line show examples of the five regions described in the text (see Sec.~\ref{s:phase}). These correspond to parameter values $A= 5$, $y_0=-1$, and $x_0=-0.90,\, 0.00,\, 0.35\, , 0.80,\, 1.62$, respectively }\label{f:phase}
\end{figure}

We now proceed to derive the equations that define the surfaces separating the different regimes which, as we will see, are accompanied by a line of critical points. For clarity, we will first give a somewhat informal discussion before addressing the problem in more detail below. Notice that Eq.~\eqref{e:f} is similar to the one yielding the equilibrium magnetization in the mean field Ising model on an external field. In analogy with the analysis of the Ising model, and following the discussion in the previous paragraph, the condition $|f^\prime(x)| = 1$ plays a central role in determining the transition between different regimes. Using Eq.~\eqref{e:dfdx}, the condition $|f^\prime(x)| = 1$ yields
\begin{equation}\label{e:Df}
\sqrt{A}\, |z|=\cosh^2\left(z^2+y_0\right),
\end{equation}
where $z=\sqrt{A}(x-x_0)$. Using Eq.~\eqref{e:Df}, rewriting the definition of $z$ as $x_0=x-{z}/{\sqrt{A}}$, and using Eq.~\eqref{e:f} to change $x$ for $f(x)$, we can write
\begin{eqnarray}
A_d(z,y_0) &=& \frac{\cosh^4\left(z^2+y_0\right)}{z^2},\label{e:Atzy0}\\
x_0(z,y_0) &=& \frac{1}{2}+\frac{1}{2}\tanh\left(z^2+y_0\right) - \frac{z}{\sqrt{A_d(z,y_0)}}.\label{e:x0zy0}
\end{eqnarray}
from which we can obtain in parametric form the surface $A_d(x_0, y_0)$ that separates the three dynamical regimes, as a function of $x_0$ and $y_0$ with parameter $z$ (Fig. \ref{fig:criticality} (a)). Fig.~\ref{f:phase} shows a level curve of this surface, which is the set of points that satisfies $A_d(x_0,y_0) = A$, with $A=5$. This value allows for a better visualization, while the discussion that follows remains qualitatively true for the case in Fig.~\ref{fig:criticality} discussed in the main text, where the parameters inferred from the experiment~\cite{garcia-lazaroEtAl2012} were used instead. 

To fix ideas before we continue with a more formal description, we first show a more graphical discussion following Fig.~\ref{f:phase} and Fig.~\ref{f:proof}. For this we fix parameters $A=5$ and $y_0 = -1$, and vary $x_0$ moving from left to right along the horizontal dashed line in Fig.~\ref{f:phase}. This figure shows five different points labeled $R_\ell$ (with $\ell=1,\dotsc 5$) on the said horizontal dashed line, which illustrate the five regions to be discussed next. Fig.~\ref{f:proof} depicts the respective functions $f(x)$ for $A=5$, $y_0 = -1$, at each of the five values of $x_0$ that correspond to the five points $R_\ell$ in Fig.~\ref{f:phase}. Notice that $f(x)$ always takes its minimum value at $x=x_0$. We have also identified four points, labeled $P_\ell$ (with $\ell=1,\dotsc 4$), where the magnitude of the slope of $f$ is exactly one, i.e., $f^\prime(P_1)=f^\prime(P_2)=1$ and $f^\prime(P_3)=f^\prime(P_4)= -1$ (see for example Fig.~\ref{f:proof}a). Starting at $R_1$ (Fig.~\ref{f:proof}a), we can then shift $f(x)$ (in red) towards the right by increasing $x_0$, and when each of the four $P_\ell$ points hit the graph of the identity function (dashed line) they become fixed points $x$ with $|f^\prime(x)|=1$ (i.e., neither stable nor unstable).  

We now describe the different ways in which the identity function can intercept the graph of $f$. Referring to  the sequence in Figs.~\ref{f:proof}(a)-(e), we can imagine that we start from $x_0 < -0.1$ (Fig.~\ref{f:proof}a) and slowly increase its value so that the function $f$ slowly moves from left to right traversing the conditions corresponding to the five points $R_\ell$ (with $\ell=1,\dotsc 5$) in Fig.~\ref{f:phase}. In this process we traverse the following five regions: 
\begin{description}
\item[Region 1] Initially $x_0$, where $f$ takes its minimum value, is negative enough to cause the graph of the identity function to intersect $f$ at a single point $x_1\approx 1$ [see Fig.~\ref{f:proof}(a)]. Since $f^\prime(x_1)\approx 0$, the fixed point $x_1$ is stable. Then, if we start increasing the value of $x_0$, the graph of $f$ will move to the right and the value of $x_1$ will decrease until the graphs of $f$ (red solid line) and the identity function (dashed line) intersect at point $P_1$ (see Fig. \ref{f:proof}a) and that situation will mark the end of Region $1$. Point $R_1$ in Fig.~\ref{f:phase} belongs to this region.
\item[Region 2] From then on, a second stable fixed point $x_2$ emerges along with an unstable fixed point $x_u$, with $x_2< x_u< x_1$ [see Fig.~\ref{f:proof}(b)]. Increasing the value of $x_0$ further, these fixed points shift to the left until $x_1$ hits point $P_2$ [see Fig.~\ref{f:proof}(b)], and then disappears. This can only happen if the curvature of the graph of $f$ is not too large. Point $R_2$ in Fig.~\ref{f:phase} belongs to this second region.
\item[Region 3] Afterwards, there is only one stable fixed point $x_2$ which shifts to the left while we keep increasing the value of $x_0$, until it hits point $P_3$ [see Fig.~\ref{f:proof}(c)], to then become unstable. This can only happen if the curvature of the graph of $f$ is not too large. Point $R_3$ in Fig.~\ref{f:phase} belongs to this region.
\item[Region 4] After this, there is only one fixed point $x_u$, which is unstable, that shifts to the left when increasing the value of $x_0$, until it hits a point $P_4$, where $f^\prime(P_4)=-1$ [see Fig.~\ref{f:proof}(d)]. Point $R_4$ in Fig.~\ref{f:phase} belongs to this region.
\item[Region 5] In this last regime, there is only one stable fixed point $x_1$ which keeps shifting to the left while we increase further the value of $x_0$ [see Fig~\ref{f:proof}(e)]. Point $R_5$ in Fig.~\ref{f:phase} belongs to this region.
\end{description}

Following the comments made in the description of regions $2$ and $3$ above, when the curvature of the graph of $f$ is large enough the order in which points $P_2$ and $P_3$ in Fig.~\ref{f:proof} meet gets inverted. However, the experimental results are not located in this regime and so we do not discuss this further. 

In Fig.~\ref{f:phase}, the level curve $A_d(x_0,y_0)=A$ defines regions inside which there are zero (blue parabolic-like area) and two (green triangular-like area) stable fixed points. These regions terminate on a critical point (red circles), where a continuous transition takes place. By varying the value of $A$, we can change those regions and the corresponding critical points, which then gives rise to lines of critical points (red dashed lines). One condition satisfied by a critical point is that there are only two points (instead of four) where the slope of $f$ has magnitude one; one of those points is the reflection of the other around $x=x_0$. This is the case if Eq.~\eqref{e:Df} has only one solution $z^\ast >0$, since this implies a second solution $-z^\ast$ by the symmetry of Eq.~\eqref{e:Df} under reflections $z\to -z$. The condition for Eq.~\eqref{e:Df} to have a unique positive solution is that the slope of the function $g(z) = \cosh^2(z^2+y_0)$ on its right hand side equals $\sqrt{A}$, i.e., 
\begin{equation}\label{e:critical}
\sqrt{A} = g^\prime(z)\equiv 4 z \sinh(z^2+y_0)\cosh(z^2+y_0).
\end{equation}
We can safely assume that $A\neq 0$ and divide Eq.~\eqref{e:critical} by Eq.~\eqref{e:Df} for $z>0$ to obtain the equation $4 z^2\tanh(z^2 + y_0) = 1$, from which we can obtain $y_{0 c}(z)$, i.e., the critical value of $y_{0}$ as a function of $z$ (see Eq.~\eqref{e:y0c} below). Knowing this, we can use Eq.~\eqref{e:Atzy0} and Eq.~\eqref{e:x0zy0} to obtain $A_c(z)$ and $x_{0 c}(z)$, i.e., the corresponding critical values of $A$ and $x_0$ as a function of $z$ (see Eqs.~\eqref{e:Ac} and~\eqref{e:x0c} below). More explicitly, the critical lines $(A_c(z), x_{0 c}(z),y_{0 c}(z))$ are described by the following equations
\begin{eqnarray}
y_{0 c}(z) &=& \tanh^{-1}\left(\frac{1}{4z^2}\right)-z^2,\label{e:y0c}\\
A_c(z) &=& \frac{\cosh^4\left[z^2+y_{0 c}(z)\right]}{z^2},\label{e:Ac}\\
x_{0 c}(z) &=& \frac{1}{2}\left(1+\frac{1}{4\, z^2}\right)-\frac{z}{\sqrt{A_c(z)}},\label{e:x0c}
\end{eqnarray}
where we have used the condition for criticality, i.e., $4 z^2\tanh(z^2 + y_0) = 1$, to obtain Eq.~\eqref{e:x0c}.

In all the discussion so far the condition $|f^\prime(x)|=1$ has played a central role. Here we show in a more detailed way why this is the case. First, notice that the function $f$ in Eq.~\eqref{e:f} essentially contains a parabola given by the expression ${A(x-x_0)^2 + y_0}$ and transforms it by applying a hyperbolic tangent, a constant scaling, and a constant offset (both equal to $1/2$) to it. The parameter $A$ defines the curvature of the parabola, while the parameters $y_0$ and $x_0$ define the minimum value it takes and where it takes it, respectively. These features remain qualitatively true for the graph of $f$, except that now $y_0$ also influences its curvature. Now, notice that the graph of the function $f$ in Eq.~\eqref{e:f} has the following properties (see Fig.~\ref{f:proof}):
\begin{description}
\item[Property 1] It is continuous and bounded, i.e., $0\leq f(x) \leq 1$ for all $x$.
\item[Property 2] It is symmetric around $x=x_0$, where it takes its minimum value, i.e., ${f(x_0-x) = f(x_0+x)}$ for ${x_0 = \arg\min_x f(x)}$.
\item[Property 3] Starting from $x=x_0$ and moving towards $x>x_0$ (respectively, $x<x_0)$, its {\em slope} monotonously increases (respectively, decreases) from zero up to a certain point, where its second derivative $f^{\prime\prime}$ vanishes, and then starts decreasing (respectively, increasing) until it asymptotically reaches zero again. In particular ${\lim_{x\to\pm\infty} f^\prime(x) = 0}$. 
\item[Property 4] By varying $x_0$ it is translated horizontally but otherwise its shape remains unchanged. In particular, this implies that $f^\prime(x)$ depends only on the difference $x-x_0$, as observed in Eq.~\eqref{e:dfdx}.
\end{description}
According to properties 1 and 4, the graph of the function $f$ in Eq.~\eqref{e:f} can intersect with the graph of the identity function in any of its points, by choosing a proper value of $x_0$ (see Fig.~\ref{f:proof}). Furthermore, due to properties 3 and 4, we can always find a value of $x_0$ for which there is at least one stable fixed point, since the function $f$ always has points with slopes as close to zero as necessary. Now, due to the continuity of $f$, if there are only two stable fixed points, say $x_1$ and $x_2 < x_1$, then there is an unstable fixed point, say $x_u$, such that $x_2 <x_u < x_1$ [see Fig.~\ref{f:proof}(b)]. In this case, because of the shape of $f$ [see Fig.~\ref{f:proof}(b)], the unstable fixed point $x_u$ and at least one of the two stable fixed points should be on the right side of $x_0$, i.e., $x_1>x_u > x_0$. Following properties 1 and 3, if there is an unstable fixed point $x_u$ such that $x_u>x_0$, there must also be two points where the slope of $f$ is equal to one; this is due to the fact that for $x>x_0$ we have $f^\prime(x) > 0$ and $f^\prime(x)$ goes to zero for both $x= x_0$ and $x\to\infty$, furthermore, $f^\prime(x_u)>1$. Hence, the existence of a point $x$ with $f^\prime(x) > 1$ signals also the existence of at least one value of the parameter $x_0$ for which there are two stable fixed points [see Fig.~\ref{f:proof}(b)]. Since $f$ is symmetric around $x=x_0$, this also implies the existence of a point $x^\prime$ such that $f(x^\prime) < -1$. The first time this happens is when a point $x$ with $|f^\prime(x)| = 1$ emerges.

\begin{figure*}\centering
\includegraphics[width=0.8\textwidth]{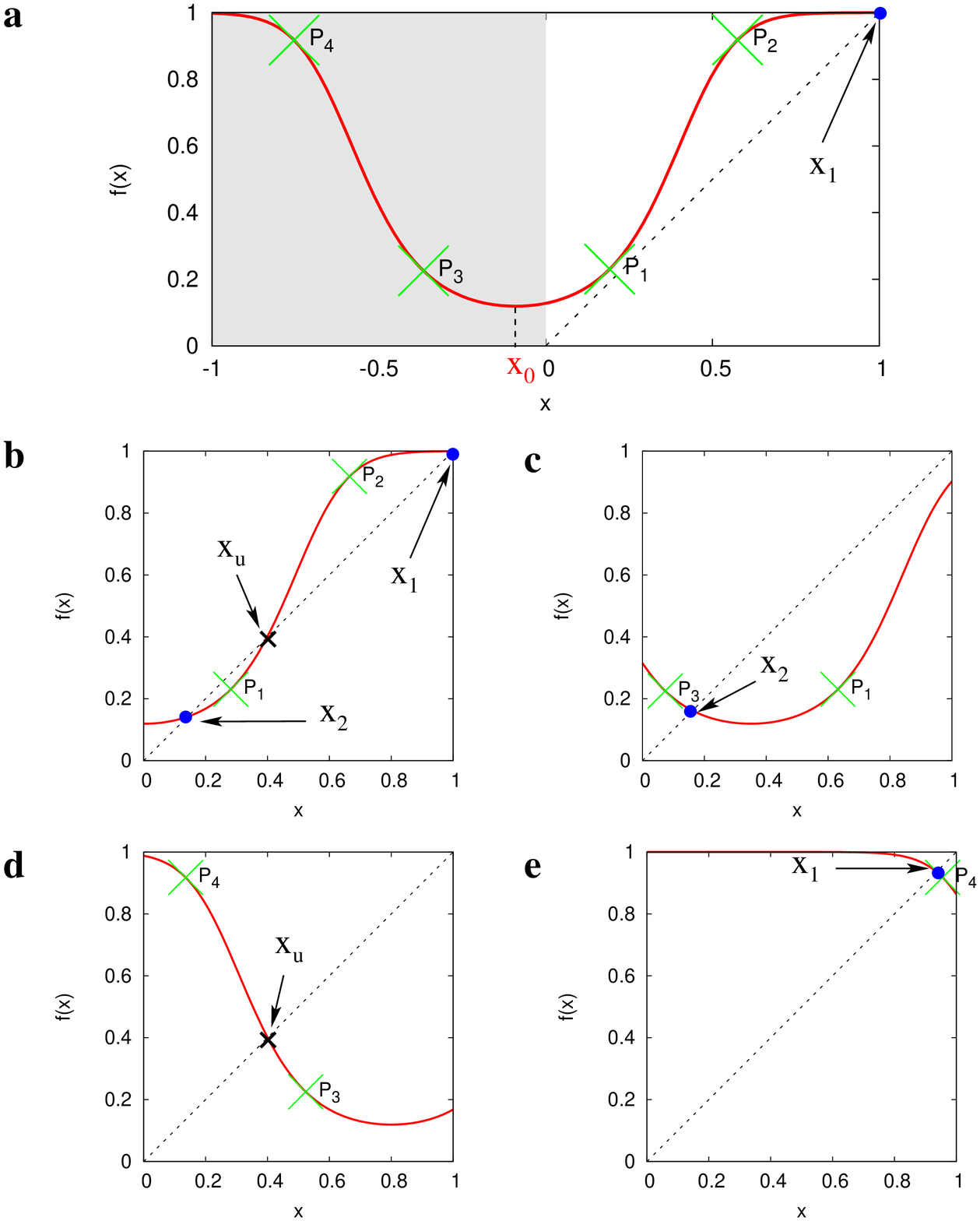}
\caption{Graph of $f$ (see Eq.~\eqref{e:f}) corresponding to points $R_1,\dotsc , R_5$ in Fig.~\ref{f:phase} (see Sec.~\ref{s:phase}). Notice that $f(P_1)=f(P_2) =1$ and $f(P_3)=f(P_4) =-1$ (green crosses). The fixed points of $f$ are its intersections with the identity function (dashed line); these are stable (blue circles) if $|f^\prime(x)| <1$ and unstable (black crosses) if $|f^\prime(x)|>1$. \textbf{a.} For small $x_0$ (e.g., point $R_1$ in Fig~\ref{f:phase}), there is only one stable fixed point $x_1$. \textbf{b.} Increasing $x_0$ (e.g., point $R_2$ in Fig.~\ref{f:phase}) until point $P_1$ touches the identity function, we enter region 2 where there are two stable fixed points $x_1$ and $x_2$, and an unstable one $x_u$ in between. \textbf{c.} Increasing $x_0$ (e.g., point $R_3$ in Fig.~\ref{f:phase}) until $x_1$ hits point $P_2$ to then disappear, we enter region 3 where only the stable fixed point $x_2$ survives. \textbf{d.} Increasing $x_0$ (e.g., point $R_4$ in Fig.~\ref{f:phase}) until $x_2$ hits the point $P_3$ to become unstable, we enter region 4 where there are no stable fixed points. \textbf{e.} Increasing $x_0$ (e.g., point $R_5$ in Fig.~\ref{f:phase}) until $x_u$ hits point $P_4$ to become stable, we enter region 5 where there is only one stable fixed point $x_1$. }\label{f:proof}
\end{figure*}

\section{\label{s:3}A diverging susceptibility}
The susceptibility $\chi$ of a system is related to its response  to a small change in the external conditions. We could ask what is the change $\delta x$ in the global level of cooperation when a generic  parameter $\theta$ of the model is varied by a small amount $\delta\theta$. We have $\delta x\approx (\partial x / \partial\theta) \delta\theta$, so $\chi = \partial x/ \partial\theta$. Arguably the most natural parameters to consider in our model are $\Delta I_C$ introduced in Eq.~\eqref{e:Idrive}, which is in principle under the influence of the experimenter, and perhaps also $h$~\cite{Rand-Nature-2012}; these two parameters influence the effective parameters $x_0$ and $y_0$ (see Eqs.~\eqref{e:x0} and~\eqref{e:y0}). Since $f$ in Eq.~\eqref{e:f} depends explicitly on the parameters $A$, $x_0$, and $y_0$ any change in a generic parameter $\theta$ that affects any of those three parameters would also affect $f$. To be more specific, let us assume that $A=A(\theta)$, $x_0=x_0(\theta)$, and $y_0=y_0(\theta)$, are well-behaved functions of $\theta$.

Deriving both sides of Eq.~\eqref{e:f} with respect to a generic parameter $\theta$ we obtain
\begin{equation}\label{e:chi}
\chi= \frac{\partial x}{\partial\theta} = f^\prime(x) \frac{\partial x}{\partial\theta} + \frac{\partial f(x)}{\partial\theta},
\end{equation}
where $f^\prime (x)$ is defined in Eq.~\eqref{e:dfdx}, and the additional term 
\begin{equation}\label{e:dfdtheta}
\frac{\partial f(x)}{\partial\theta} = \frac{\partial f(x)}{\partial A} \frac{\partial A(\theta)}{\partial\theta} + \frac{\partial f(x)}{\partial x_0} \frac{\partial x_0(\theta)}{\partial\theta} + \frac{\partial f(x)}{\partial y_0} \frac{\partial y_0(\theta)}{\partial\theta},
\end{equation}
takes into account the explicit dependence of $f$ on the parameters of the model, which vary when varying $\theta$. The term defined in Eq.~\eqref{e:dfdtheta} is smooth as long as we can assume, as we do here, that there are no spurious singularities in the definition of $A(\theta)$, $x_0(\theta)$, and $y_0(\theta)$.

Solving Eq.~\eqref{e:chi} for the susceptibility $\chi=\partial x/\partial\theta$ we get
\begin{equation}\label{e:chi2}
\chi = \frac{1}{1-f^\prime(x)} \frac{\partial f(x)}{\partial\theta},
\end{equation}
which clearly diverges when, by varying $\theta$, the fixed point $x$ under consideration \textit{crosses} continuously a point $x_c$ where $f^\prime(x_c) = 1$, i.e., a critical point. For illustration purposes, let us assume that the model parameters vary as $A(\theta) = A_c^\ast - \theta $, $x_0(\theta)=x_{0 c}^\ast$, and $y_0(\theta)=y_{0 c}^\ast$, for a generic parameter $\theta\geq 0$; here $A_c^\ast = A_c(z^\ast)$, $x_{0 c}^\ast = x_{0 c}(z^\ast)$, and $y_{0 c}^\ast = y_{0 c}(z^\ast)$ correspond to a point on the critical line specified by a particular value of $z^\ast$ through Eqs.~\eqref{e:y0c},~\eqref{e:Ac},~and~\eqref{e:x0c}. Using Eqs.~\eqref{e:dfdx} and~\eqref{e:Ac}, we can write $f^\prime(x) = \sqrt{A(\theta) / A_c^\ast}$,  which for small $\theta$ can be approximated as $f^\prime(x)\approx 1 - \theta/2A_c^\ast$. So, following Eq.~\eqref{e:chi2}, when approaching the critical point, i.e., $\theta\to 0$ or $A\to A_c^\ast$, the susceptibility diverges as 
\begin{equation}
\chi\propto \frac{1}{\theta} = \frac{1}{A_c^\ast - A}\:\xrightarrow[A\to A_c^\ast]{}\:\infty, 
\end{equation}
where in the last expression we have written $\theta =A_c^\ast - A$ in terms of $A$. The divergence of the susceptibility is one of the hallmarks of criticality~\cite{Mora-StatPhys-2011}.
%

\section{\label{s:cExp}Connection to experiments}
\subsection{\label{s:mcc}Moody conditional cooperation}
As mentioned in the main text, experiments show that the probability for a human to cooperate in a generic round of the game depends on whether she cooperated or not and how many of her peers cooperated. Here we explain how to connect this so-called moody conditional cooperation (MCC) rule with the mean field model described by Eq.~\eqref{eq:dynamic}. 

Indeed, the MCC rule can be expressed mathematically in terms of the conditional probability ${P(C;t+1|s,n;t)}$ for a generic agent to cooperate (C) at round ${t+1}$ given that she played strategy $s$ and that $n$ of her peers cooperated at round $t$. More precisely, the probability which the MCC rule refers to can be written as ${P(C|s,n) = (1/T)\sum_t P(C;t+1|s,n;t)}$, where $T$ is the total number of rounds. We assume, however, that $T$ is sufficiently large and that the system reaches a stationary state. In this case, the MCC rule is given by the conditional probability corresponding to the stationary state ($t\to\infty$), and we can drop the index $t$; we will keep the dependence on $t$ for the most part to facilitate the discussion, though. We  further assume that the stationary state can be described by the long term dynamics of the mean field model. 

Depending on the context, we will use interchangeably $s=$C or $s=1$ to refer to cooperation, and similarly we will use interchangeably $s=D$ or $s=0$ to refer to defection.

Now, writing $\Delta U_i(t) = \Delta U_i(s_i , n_i)$ (see Eq.~\eqref{e:dUsn}), the right hand side of Eq.~\eqref{e:x-update} gives the probability that an agent $i$ cooperates at round $t+1$ given that, at round $t$, she played strategy $s_i$, that $n_i=\sum_{j\in \partial i} s_j$ of her peers cooperated, and that her cooperation probability was $x_i$. Indeed this is a more detailed reading of Eq.~\eqref{e:x-update}. Since in the mean field approximation we are interested in a representative agent, we can drop the indexes and write
\begin{equation}\label{e:input}
P(C;t+1|s,n,x;t)=\frac{x^{1-\alpha}}{x^{1-\alpha}+\left(1-x\right)^{1-\alpha}e^{-\beta \Delta U(s,n)}}, 
\end{equation}
where $\Delta U(s,n) = (a s + b ) n + 2 h s - h$ (see Eq.~\eqref{e:dUsn}).

This conditional probability distribution depends on $x$, while the MCC rule does not. Informally, if we assume that the system is mono-stable we can get rid of the dependence on $x$ to obtain
\begin{equation}\label{e:Pmain}
P(C|s,n)=\frac{1}{1+y_1^{1-\alpha}e^{-\beta \Delta U(s,n)}}, 
\end{equation}
where $y_1 = (1-x_1)/x_1$, and $x_1$ is the only stable fixed point; Eq.~\eqref{e:Pmain} coincides with Eq.~\eqref{e:dut} in the main text. 

In the following we use the rules of probability theory to obtain a more general form of $P(C|s,n)$ from Eq.~\eqref{e:input} that reduces to Eq.~\eqref{e:Pmain} if we assume that the system is mono-stable. Let us first write
\begin{equation}\label{e:bayes}
P(C;t+1|s,n;t) = \frac{\int_0^1 P\left(\{C; t+1\},\{s,n,x; t\}\right)\mathrm{d} x}{\int_0^1 P(s,n,x; t)\mathrm{d}x},
\end{equation}
by definition of conditional probability. The term inside the integral in the numerator is the joint probability of all the variables involved. We have emphasized between brackets which round the variables refer to. On the other hand, the term inside the integral in the denominator is the joint probability of all the variables that refer only to round $t$.

Using the chain rule of probability theory we can express such joint probabilities as
\begin{eqnarray}
P\left(\{C;t+1\},\{s,n,x;t\}\right) &=& P(C;t+1|s,n,x;t)\times\nonumber\\ 
 &\times& P(s,n;t|x;t)P(x;t),\label{e:one}\\
P(s,n,x; t) &=& P(s,n;t|x;t)P(x;t),\label{e:two}
\end{eqnarray}

Let $N$ denote the set of neighbors of the representative agent, and let $s_N=\{s_j | j\in N\}$ and $x_N=\{x_j | j\in N\}$ be the set of their strategies and cooperation probabilities, respectively. Then 
\begin{equation}\label{e:binomial}
P(s,n;\, t|x;\, t)= \sum_{s_N} P(s,s_N;\, t|x;\, t)\,\delta\left[n = \sum_{j\in N} s_j\right],
\end{equation}
where $\delta[p]$ is the indicator function, which is equal to one if proposition $p$ is true and zero otherwise. Here $P(s,s_N;\, t|x;\, t)$ is the conditional probability that the representative agent play strategy $s$ and her peers play strategies $s_N$, \textit{jointly}, at round $t$ given that the probability for the representative agent to cooperate at the same round is $x$. The definition of conditional probability and the chain rule allow us to write
\begin{widetext}
\begin{equation}\label{e:PssNx}
P(s,s_N;\, t|x;\, t)= \frac{\int_0^1 P(s,s_N;\, t|x, x_N;\, t) P(x,x_N;\, t)\,\prod_{j\in N}\mathrm{d}x_j}{P(x;\, t)},
\end{equation}
\end{widetext}
where the term inside the integral in the numerator is the joint probability that at round $t$ the representative agent and her neighbors play strategies $s$ and $s_N$ and their cooperation probabilities are $x$ and $x_N$, respectively. The integral marginalizes this probability over $x_N$ leaving the joint probability of the variables $s$, $s_N$, and $x$. 

Although the next sentence may be redundant, its sole intention is to put everything in the formalism we are describing here: The probability that an agent plays strategy $s$ at round $t$, given that the probability to cooperate at the same round $t$ is $x$, can be written as $P(s;\, t|x;\, t)= x^s(1-x)^{1-s}$. Furthermore, at each round each agent picks her strategy independently of the rest. So
\begin{equation}
\begin{split}
P(s,s_N;\, t|x, x_N;\, t)=P(s;\, t|x;\, t)\prod_{j\in N} P(s_j;\, t|x_j;\, t)\\ =x^s(1-x)^{1-s}\prod_{j\in N} x_j^{s_j}(1-x_j)^{1-s_j}.
\end{split}
\end{equation}
This reflects the fact that correlations in the system come only through the correlations in the cooperation probability accumulated during the history of play.
Equation (\ref{e:PssNx}) can then be written as
\begin{widetext}
\begin{equation}
P(s,s_N;\, t|x;\, t)=\\ \frac{x^s(1-x)^{1-s}\,\int_0^1 P(x, x_N;\, t)\prod_{j\in N} x_j^{s_j}(1-x_j)^{1-s_j}\, \mathrm{d}x_j}{P(x;\, t)}.
\end{equation}
\end{widetext}
Only now we resort to the mean field approximation which neglects correlations altogether to write $P(x,x_N;\, t)\approx P(x;\, t)\prod_{j\in N} P(x_j;\, t)$.  Since we are interested in the stationary state, we can drop the round index $t$ in the expressions that follow. So
\begin{equation}\label{e:pair}
P(s,s_N;\, t|x;\, t)=  x^s(1-x)^{1-s}\,\rho^{s_j}(1-\rho)^{1-s_j},
\end{equation}
where $\rho=\int_0^1 x_j P(x_j)\mathrm{d}x_j$ is the average probability that a neighbor cooperates, which equals the average probability that the representative agent cooperates since we are working within a mean field approximation. Introducing Eq.~\eqref{e:pair} into Eq.~\eqref{e:binomial}, we get
\begin{equation}\label{e:combinatorial}
P(s,n,t|x,t)=  x^s(1-x)^{1-s}\,\binom{K}{n} \rho^{n}(1-\rho)^{K-n},
\end{equation}
where $\binom{K}{n}=K!/n!(K-n)!$ is the binomial coefficient. Introducing Eqs.~\eqref{e:combinatorial}~and~\eqref{e:input} into Eqs.~\eqref{e:bayes} and~\eqref{e:one}, Eq.~\eqref{e:two} yields the desired result
\begin{widetext}
\begin{equation}\label{e:mcc-mf-full}
\begin{split}
P(C,t+1|s,n,t) = \frac{1}{\rho^s(1-\rho)^{1-s}} \int_0^1 \frac{x^{s+1-\alpha}(1-x)^{1-s}\, P(x)}{x^{1-\alpha}+\left({1-x}\right)^{1-\alpha}e^{-\beta \Delta U(s,n)}}\mathrm{d} x,
\end{split}
\end{equation}
\end{widetext}
where we have used $\int_0^1 x^s(1-x)^{1-s} \mathrm{d} x = \rho^s (1-\rho)^{1-s}$. Using the change of variables $y=(1-x)/x$, which is monotonous for $x\in (0,1)$, we have
\begin{widetext}
\begin{equation}\label{e:Py}
P(C,t+1|s,n,t) = \frac{1}{\rho^s(1-\rho)^{1-s}}\int_0^\infty \frac{y^{1-s}(1+y)^{-1}\, P_Y\left(y\right)}{1+y^{1-\alpha}e^{-\beta \Delta U(s,n)}}\mathrm{d} y.
\end{equation}
\end{widetext}

Since we have assumed that $\beta$ is small, we can expand the right hand side of Eq.~\eqref{e:Py} to first order in $\beta$ to obtain
\begin{equation}\label{e:gmcc}
P(C,t+1|s,n,t) = I_s(\alpha)+\beta J_s(\alpha) \Delta U(s,n),
\end{equation}
where
\begin{widetext}
\begin{eqnarray}
I_s(\alpha) &=&  \frac{1}{\rho^s(1-\rho)^{1-s}}\int_0^\infty \frac{y^{1-s}(1+y)^{-1}\, P_Y\left({y}\right)}{1+y^{1-\alpha}}\mathrm{d} y, \label{e:Is} \\
J_s(\alpha) &=& \frac{1}{\rho^s(1-\rho)^{1-s}}\int_0^\infty \frac{y^{2-\alpha-s}(1+y)^{-1}\, P_Y\left({y}\right)}{(1+y^{1-\alpha})^2}\mathrm{d} y.\label{e:Js}
\end{eqnarray}
\end{widetext}

Using Eq. (\ref{e:dUsn}), we can see that Eq.~\eqref{e:gmcc} is of the form
\begin{equation}\label{e:mcc}
P(C,t+1|s,n,t) = m_s n/ K + r_s,
\end{equation}
where 
\begin{eqnarray}
m_s &=& \beta K J_s(\alpha)(a s + b),\label{e:mcc-1fpA}\\
r_s &=& I_s(\alpha)+\beta J_s(\alpha)\left[h(2 s- 1) \right]\label{e:mcc-1fpB}.
\end{eqnarray}

This yields the general expression introduced in the main text paragraph after Eq.~\eqref{e:dut}; in the next section, we make use of the assumption of mono-stability.

\subsection{\label{s:mcc_mono}Regime of mono-stability}
We now make use of the assumption that the long-term dynamics of the system is well described by the stable fixed points of the mean field dynamics. So, ${P(x)=(1-\mu)\delta (x-x_1)+\mu\delta(x-x_2)}$, where $x_1$ and $x_2$ are the fixed points of Eq.~\eqref{eq:dynamic} and $\mu$ yields their corresponding weights. In this case, the average probability for the representative agent to cooperate is $\rho=(1-\mu) x_1 + \mu x_2$. If there is only one fixed point we can take $\mu = 0$. If there are no fixed points, the analysis here does not apply. For simplicity, and in agreement with the experiment we analyze~\cite{garcia-lazaroEtAl2012}, we assume that the dynamics is essentially mono-stable, say $\mu\approx 0$. This implies that the experimental global average cooperation level in the stationary state $\bar{x}$ is close to the relevant fixed point, i.e., $\bar{x}\approx x_1$.

In this case we have $\rho = x_1$ and $P_Y(y)=\delta(y-y_1)$ with $y_1=(1-x_1)/x_1$, which yields
\begin{equation}\label{e:mcc-mf-delta}
P(C;\, t+1|s,n;\, t) = \frac{1}{1+y_1^{1-\alpha}e^{-\beta \Delta U(s,n)}};
\end{equation}
this is the expression in Eq.~\eqref{eq:prob} in the main text. In this case both terms in Eqs.~\eqref{e:Is} and~\eqref{e:Js} become
\begin{eqnarray}
I_s(\alpha) &=& I(\alpha) \equiv \frac{1}{1+y_1^{1-\alpha}},\\
J_s(\alpha) &=& J(\alpha) \equiv  \frac{y_1^{1-\alpha}}{\left(1+y_1^{1-\alpha}\right)^2},
\end{eqnarray}
independent of $s$. Eqs.~\eqref{e:mcc-1fpA} and~\eqref{e:mcc-1fpB} are better described in terms of the following quantities
\begin{eqnarray}
r &=& \frac{1}{2}(r_C+r_D) = I(\alpha),\label{e:r}\\
G &=& r_C - r_D = 2\beta h J(\alpha).\label{e:G}
\end{eqnarray}
Here $r$ and $G$ are the mean intercept and the `gap' between intercepts of the near linear trends that describe the MCC rule~\cite{garcia-lazaroEtAl2012}, respectively. We can safely assume that $b\neq 0$, which gives
\begin{eqnarray}
m_C-m_D &=& \beta\, a K J(\alpha),\label{e:dm}\\
\frac{m_C}{m_D} &=& \frac{\beta\, a+\beta\, b}{\beta\, b}.\label{e:mCmD}
\end{eqnarray}
Now, Eq.~\eqref{e:r} can be readily inverted to obtain $\alpha$ in terms of the experimental quantity $r$ [see Eq.~\eqref{e:alphaexp} below]. Similarly, using  Eq.~\eqref{e:r} we can write ${J(\alpha) = I(\alpha)[1-I(\alpha)] = r(1-r)}$, and so Eq.~\eqref{e:G} can be readily inverted to obtain $\beta h$ in terms of the experimental quantities $G$ and $r$ (see Eq.~\eqref{e:hexp0} below). Finally, Eqs.~\eqref{e:dm} and~\eqref{e:mCmD} can be inverted to obtain $\beta a$ and $\beta b$ in terms of the experimental quantities $m_C$, $m_D$, $r$, and $K$ [see Eqs.~\eqref{e:aexp0} and~\eqref{e:bexp0} below]. This yields
\begin{eqnarray}
\alpha &=& 1-\frac{\log\left[(1-r)/{r}\right]}{\log\left[{(1-\bar{x})}/{\bar{x}}\right]},\label{e:alphaexp}\\
\beta h &=& \frac{G}{2 r (1-r)},\label{e:hexp0}\\
\beta a &=& \frac{m_C-m_D}{K r (1-r)},\label{e:aexp0}\\
\beta b &=& \frac{m_D}{K r (1-r)},\label{e:bexp0}
\end{eqnarray}
where we have used the condition that the only stable fixed point should equal the experimental global cooperation level, i.e., $x_1=\bar{x}$, to obtain Eq.~\eqref{e:alphaexp}.

Although these equations leave the parameter $\beta$ undetermined, this combination of parameters completely determines the coefficients that define the mean field dynamics through Eq.~\eqref{eq:dynamic}, and the parameters $A$, $x_0$ and $y_0$ that locate the system in the phase diagram. Indeed, multiplying by $\beta$ in the numerator and denominator of Eqs.~\eqref{e:A},~\eqref{e:x0}, and~\eqref{e:y0}, using Eqs.~\eqref{e:alphaexp},~\eqref{e:hexp0},~\eqref{e:aexp0}, and~\eqref{e:bexp0}, and doing some algebra we obtain the expressions
\begin{eqnarray}
A &=& \frac{m_C-m_D}{2\, \alpha\, r\, (1-r)},\label{e:Ainf} \\
x_0 &=& -\frac{m_D+G}{2(m_C-m_D)},\label{e:x0inf} \\
y_0 &=& -\frac{m_D^2 + 2m_C G+ G^2}{8\, \alpha\, r\, (1-r) (m_C - m_D)}, \label{e:y0inf}
\end{eqnarray}
where the expression for $\alpha$ is given in Eq.~\eqref{e:alphaexp}. Eqs.~\eqref{e:Ainf},~\eqref{e:x0inf}, and~\eqref{e:y0inf} in principle allow us to locate the system in the phase diagram of Fig. $1$a of the main text. However, we still need to check that these parameter values produce a dynamics through Eq.~\eqref{eq:dynamic} that indeed agree with the dynamics observed in experiments, within the margin of error of the experimental results. Furthermore, we should check that indeed the assumption of mono-stability is indeed satisfied, i.e., that $x_1\approx\bar{x}$. 

Before finishing this section notice that we can invert Eqs.~\eqref{e:A},~\eqref{e:x0}, and~\eqref{e:y0} to recover the parameters defining the mean field dynamics in Eq.~\eqref{eq:dynamic}, which yields
\begin{eqnarray}
\beta a &=& \frac{2\alpha A}{K},\label{e:aexp}\\
\beta b &=& \frac{4\alpha}{K}\left[ y_0 - x_0 (1-x_0) A\right],\label{e:bexp}\\
\beta h &=& -2\alpha\left(y_0+x_0^2 A\right); \label{e:hexp}
\end{eqnarray}
here we made use of the expression $y_0=-(\tilde{a} x_0^2+h)/2\gamma$. 

\subsection{\label{s:mcc2}Regime of bi-stability}
In case the system is bi-stable with a non-negligible value of $\mu$, we have to deal with the whole expression ${P(x)=(1-\mu)\delta (x-x_1)+\mu\delta(x-x_2)}$ and Eqs.~\eqref{e:Is} and~\eqref{e:Js} become
\begin{widetext}
\begin{eqnarray}
I_s(\alpha) &=& W_{1,s} \Upsilon_1(\alpha)+W_{2,s} \Upsilon_2(\alpha), \\
J_s(\alpha) &=& W_{1,s} \Upsilon_1(\alpha)[1-\Upsilon_1(\alpha)] + W_{2,s} \Upsilon_2(\alpha)[1-\Upsilon_2(\alpha)],
\end{eqnarray}
\end{widetext}
where we have defined the expressions
\begin{eqnarray}
\Upsilon_\ell &=& \frac{1}{1+y_\ell^{1-\alpha}},\\
W_{\ell, s} &=& \mu^{i-1}(1-\mu)^{2-i}\left(\frac{x_i}{\rho}\right)^s \left(\frac{1-x_i}{1-\rho}\right)^{1-s}.
\end{eqnarray}

Eqs.~\eqref{e:mcc-1fpA} and~\eqref{e:mcc-1fpB} defining the slopes and intercepts can still be inverted to obtain
\begin{eqnarray}
\beta a K &=& \frac{m_C}{J_1(\alpha)} - \frac{m_D}{J_0(\alpha)},\label{e:a2fp}\\
\beta b K &=& \frac{m_D}{J_0(\alpha)},\label{e:b2fp}\\
\beta h &=& \frac{1}{2}\left[\frac{r_C}{J_1(\alpha)} - \frac{r_D}{J_0(\alpha)} + \frac{I_0(\alpha)}{J_0(\alpha)} - \frac{I_1(\alpha)}{J_1(\alpha)}\right].\label{e:h2fp}
\end{eqnarray}
However, the corresponding values of $\alpha$ are given implicitly by solutions to the equation
\begin{equation}\label{e:alpha2fp}
J_0(\alpha)[r_C-I_1(\alpha)] + J_1(\alpha)[r_D-I_0(\alpha)] = 0.
\end{equation}

In the bi-stable regime, we can parametrize the model in terms of $\alpha$, the two stable fixed points, $x_1$ and $x_2$, and the unstable fixed point $x_u$. Indeed, if we write the fixed point equation~\eqref{e:f} as $A (x-x_0)^2 + y_0 = \tanh^{-1}(2 x-1)$, evaluate it at two of the fixed points, say $x_1$ and $x_2$, and subtract the two corresponding equations, we obtain the following expression for $A$
\begin{equation}\label{e:Ax0x1x2}
A = F_A(x_2, x_1|x_0) \equiv \frac{\Delta(x_2 , x_1)}{(x_2-x_1)(x_2+x_1 - 2 x_0)},
\end{equation}
which defines the function $F_A(x_2,x_1|x_0)$; here
\begin{equation}
\Delta(x_2, x_1) \equiv \tanh^{-1}(2 x_2- 1) - \tanh^{-1}(2 x_1- 1).
\end{equation}
Any two fixed points that we chose would give us different expression for $A$, which should be consistent. In particular, we should have $A=F_A(x_2,x_u|x_0) = F(x_u,x_1|x_0)$, from which we can obtain an expression for $x_0$ in terms of the three fixed points, i.e.,
\begin{equation}\label{e:x0xux1x2}
x_0 = \frac{(x_2^2-x_u^2)\Delta(x_u,x_1) - (x_u^2-x_1^2)\Delta(x_2,x_u)}{(x_2-x_u)\Delta(x_u,x_1) - (x_u-x_1)\Delta(x_2,x_u)}.
\end{equation}
So, we can introduce any feasible values for $x_1$, $x_2$, and $x_u$ into Eq.~\eqref{e:x0xux1x2} to obtain the corresponding value for $x_0$; from Eq.~\eqref{e:Ax0x1x2} we can obtain the corresponding value for $A$, and from Eq.~\eqref{e:f} the corresponding value for $y_0$.

\section{\label{s:inference}From experimental data to model parameters}

\ 

\begin{table}\centering
\begin{tabular}{l   r   r   r   r}
\hline
Quantity \indent  & \indent Square lattice \indent & \indent Heterogeneous network  \\
\hline
$m_C$  & $0.122\pm 0.034$ &  $0.126 \pm 0.039$   \\
$m_D$  & $-0.149\pm 0.050$ &  $-0.0269\pm 0.035$  \\
$r_C$    & $0.457\pm 0.015$ &  $0.475\pm 0.016$  \\
$r_D$    & $0.350\pm 0.021$ &  $0.309\pm 0.069$  \\
$\bar{x}$  & $0.306\pm 0.024$ &  $0.355\pm 0.021$\\
\hline
\end{tabular}
\caption{\label{t:data}Experimental data for the two experiments performed in Zaragoza reported in Ref.~\cite{garcia-lazaroEtAl2012}, which were carried out on a square lattice and on a heterogeneous network. The first four rows are extracted from Table~S2 in Ref.~\cite{garcia-lazaroEtAl2012} and correspond to a linear fit of Figs.~3A,~B in Ref.~\cite{garcia-lazaroEtAl2012}, while the last row is obtained from averaging the last ten rounds in Fig. 2A in Ref.~\cite{garcia-lazaroEtAl2012} (see Sec.~\ref{s:mono} above).}
\end{table}

\begin{table*}[ht]
\centering
\begin{tabular}{l r r}
\hline
Parameter & Heterogeneous network & Square lattice\\
\hline
$\widehat{m}_C$ & $0.0545$ & $0.0868$        \\
$\widehat{m}_D$ & $-0.0901$ & $-0.2095$      \\
$\widehat{r}$ & $0.3917$  & $0.38765$    \\
$\widehat{G}$ & $0.161$  & $0.1809$    \\
$\widehat{x}(0)$ & $0.5328$  & $0.580$\\
\hline
\end{tabular}

{\begin{tabular}{rrrrr}
  \hline
 \multicolumn{5}{c}{Covariance matrix (Heterogeneous network)} \\ 
  \hline
 $1.3 \times 10^{-5}$ & $-7.3 \times 10^{-6}$ & $8.5 \times 10^{-8}$ & $4.3 \times 10^{-6}$ & $-3.8 \times 10^{-6}$ \\ 
  $-7.3 \times 10^{-6}$ & $1.3 \times 10^{-5}$ & $-1.8 \times 10^{-6}$ & $6.1 \times 10^{-6}$ & $-3.6 \times 10^{-6}$ \\ 
  $8.5 \times 10^{-8}$ & $-1.8 \times 10^{-6}$ & $7.1 \times 10^{-5}$ & $4.9 \times 10^{-4}$ & $-1.0 \times 10^{-6}$ \\ 
 $4.3 \times 10^{-6}$ & $6.1 \times 10^{-6}$ & $4.9 \times 10^{-4}$ & $3.5 \times 10^{-3}$ & $-2.0 \times 10^{-5}$ \\ 
  $-3.8 \times 10^{-6}$ & $-3.6 \times 10^{-6}$ & $-1.0 \times 10^{-6}$ & $-2.0 \times 10^{-5}$ & $2.2 \times 10^{-5}$ \\ 
   \hline
\end{tabular}
\begin{tabular}{rrrrr}
  \hline
 \multicolumn{5}{c}{Covariance matrix (Square lattice)} \\ 
  \hline
 $3.4 \times 10^{-5}$ & $-5.5 \times 10^{-6}$ & $-7.6 \times 10^{-7}$ & $3.4 \times 10^{-6}$ & $-3.0 \times 10^{-5}$ \\ 
 $-5.5 \times 10^{-6}$ & $8.7 \times 10^{-6}$ & $-3.3 \times 10^{-7}$ & $1.1 \times 10^{-6}$ & $-5.4 \times 10^{-6}$ \\ 
 $-7.6 \times 10^{-7}$ & $-3.3 \times 10^{-7}$ & $4.1 \times 10^{-7}$ & $9.1 \times 10^{-8}$ & $6.7 \times 10^{-7}$ \\ 
  $3.4 \times 10^{-6}$ & $1.1 \times 10^{-6}$ & $9.1 \times 10^{-8}$ & $1.0 \times 10^{-5}$ & $-4.3 \times 10^{-6}$ \\ 
  $-3.0 \times 10^{-5}$ & $-5.4 \times 10^{-6}$ & $6.7 \times 10^{-7}$ & $-4.3 \times 10^{-6}$ & $3.2 \times 10^{-4}$ \\ 
   \hline
\end{tabular}}
\caption{\label{t:ZH} Parameters inferred for the experiment~\cite{garcia-lazaroEtAl2012} on a heterogeneous network and a square lattice using $\zeta = 1.96$ and $\zeta = 1.28$, respectively (see Sec.~\ref{s:mono}).}
\end{table*}

\subsection{\label{s:general}General considerations}
Here we depict how we estimated the parameters of the model described by Eq.~\eqref{eq:dynamic}. We follow the framework of partially observed Markov processes~\cite{pomp} and assume that the system is described by an underlying deterministic dynamics $x(t)$ satisfying Eq.~\eqref{eq:dynamic}. We further assume that the scientist in the laboratory observes a noisy version $x_{\rm obs}(t)$ of the underlying dynamics characterized by a probability distribution $\mathcal{P}_{\rm obs}[x_{\rm obs}(t)|x(t)]$. This collectively describes experimental uncertainty as well as intrinsic stochasticity~\cite{galla2009,Realpe-JSTAT-2012} in the system dynamics that has been averaged out in the derivation of Eq.~\eqref{eq:dynamic}.

To each observed trajectory $\mathbf{x}_{\rm obs}(1:T)\equiv\{x(t) : t=1,\dotsc ,T\}$ of $T$ rounds, and a trajectory of the underlying dynamics $\mathbf{x}(0:T)\equiv\{x(t) : t=0,\dotsc ,T\}$, we can assign a joint probability distribution
\begin{widetext}
\begin{equation}\label{e:PbayesJoint}
\mathcal{P}[\mathbf{x}(0:T), \mathbf{x}_{\rm obs}(1:T)|\Theta] = \mathcal{P}_0[x(0)]\,\prod_{t=1}^T\,\mathcal{P}_{\rm obs}[x_{\rm obs}(t)|x(t)]\,\mathcal{P}_{\rm dyn}[x(t)|x(t-1)|\Theta],
\end{equation}
\end{widetext}
where $\Theta$ represents the parameters that define the system dynamics via Eq.~\eqref{eq:dynamic}, $\mathcal{P}_0[x(0)]$ represents the probability distribution over the initial condition $x(0)$ of the underlying dynamics, and $\mathcal{P}_{\rm dyn}[x(t)|x(t-1)|\Theta]$ is a Dirac delta function representing the deterministic dynamics described by Eq.~\eqref{eq:dynamic}. Finally, we assume that ${\mathcal{P}_{\rm obs}[x_{\rm obs}(t)|x(t)] = \mathcal{N}[x_{\rm obs}(t); x(t),\sigma]}$ is a Gaussian distribution of mean $x(t)$ and standard deviation $\sigma$ to be determined later.

From the joint  distribution defined in Eq.~\eqref{e:PbayesJoint} we can obtain the probability or likelihood of observing a particular realization of the observed dynamics $\mathcal{P}_{\rm lik}[\mathbf{x}_{\rm obs}(1:T)|\Theta]$ given the parameters $\Theta$. Our aim is to determine the probability of the parameters given a particular realization of the observed dynamics which, following Bayes' rule, is given by  ${\mathcal{P}_{\theta}[\Theta |\mathbf{x}_{\rm obs}(1:T)]\propto \mathcal{P}[\mathbf{x}_{\rm obs}(1:T)|\Theta]\mathcal{P}_{\rm prior}[\Theta]}$, where $\mathcal{P}_{\rm prior}[\Theta]$ contains the prior information on the parameters $\Theta$. We estimate the parameter values by the average over the posterior, i.e., ${\widehat{\Theta} = \int\Theta\,\mathcal{P}_{\theta}[\Theta |\mathbf{x}_{\rm obs}(1:T)]\,\mathrm{d}\Theta}$. 

What follows depends on whether we are in a regime of mono-stability or bi-stability. 
	
\subsection{\label{s:mono}Parameter estimation in the regime of mono-stability}
We build the prior $\mathcal{P}_{\theta}(\Theta)$ in an indirect way. First, we notice that Eqs.~\eqref{e:alphaexp},~\eqref{e:hexp0},~\eqref{e:aexp0}, and~\eqref{e:bexp0} yield a set of values for the parameters that, along with the initial condition $x{(0)}$, define the system dynamics through Eq.~\eqref{eq:dynamic}. So we can parametrize our model in Eq.~\eqref{eq:dynamic} by the {\it collection} of experimental observables $\Theta =\mathcal{O} \equiv (m_C, m_D, r_C, r_D)$. To deal with the uncertainty in the experimental results, we assume that any observable $O$ reported in the literature is described by a uniform probability distribution supported in the interval ${[{O}^\ast-\zeta\delta O^\ast, {O}^\ast+\zeta\delta O^\ast]}$. Here ${O}^\ast$ and $\delta O^\ast$ are, respectively, the value reported for $O$ and its corresponding standard error (Tab.~\ref{t:data}); the parameter $\zeta$ is used to define a credible interval of the reported experimental results, e.g., if $\zeta \approx 1.28$ or $\zeta \approx 1.96$ we are dealing with a $90\%$ or $97.5$ credible interval, respectively. We rely on a uniform rather than a Gaussian distribution on experimental results to avoid the statistics be dominated by rare events (e.g., due to the logarithmic term in Eq.~\eqref{e:alphaexp}). We will use this as a prior distribution $\mathcal{P}(\mathcal{O})$. To compute $\alpha$ from $\mathcal{O}$ via Eq.~\eqref{e:alphaexp}, we assume that the fixed point equals the average global cooperation over the last ten rounds of the experiment, i.e., $x_1\approx \bar{x}$. In this way we avoid the technical difficulty that the prior would actually depend on the final state of the dynamics, i.e., the fixed point. We also estimate the standard deviation $\sigma$ of the observation error to be equal to the standard deviation of the last ten points in the time series $\mathbf{x}_{\rm obs}(1:T)$. We assign, however, a standard deviation three times larger $3\sigma$ to the first two points in the dynamics to take into account that the adiabatic approximation is expected to capture better the slower dynamics that follows the initial transient regime of rather fast decay.

\subsection{\label{s:bi}Parameter estimation in the regime of bi-stability}
This case is a bit more complex since now the relationships between parameters and experimental values, i.e., Eqs.~\eqref{e:a2fp},~\eqref{e:b2fp},~\eqref{e:h2fp},~and~\eqref{e:alpha2fp}, depend non-trivially on the fixed points of the underlying dynamics. Moreover, in contrast to the previous case, here we cannot disentangle this dependence as the observed long-term cooperation level is related to the underlying dynamics by $\bar{x} = (1-\mu)\, x_1^\ast + \mu\, x_2^\ast$. The prior here is also defined indirectly. 

As discussed in Sec.~\ref{s:mcc2}, in this regime it is convenient to parametrize the model in Eq.~\eqref{eq:dynamic} in terms of the two stable fixed points, $x_1^\ast$ and $x_2^\ast$, the only unstable fixed point, $x_u^\ast$, fixed point and $\alpha$, i.e., $\Theta = (\alpha, x_1^\ast, x_2^\ast, x_u^\ast)$. To take into account the influence of the two stable fixed points on the underlying dynamics, we describe the observed dynamics as $x_{\rm obs}(t) = (1-\mu)\, x_1(t) + \mu \, x_2(t)$, where $x_1(t)$ and $x_2(t)$ represents the dynamics given by Eq.~\eqref{eq:dynamic} with two different initial conditions $x_1(0)$ and $x_2(0)$. For any given choice of the parameters $\Theta$, we use Eqs.~\eqref{e:Ax0x1x2}~and~\eqref{e:x0xux1x2}, and invert Eq.~\eqref{e:f} to compute the corresponding values of $A$, $x_0$, and $y_0$. With these and the parameter $\alpha$, we can use Eqs.~\eqref{e:aexp},~\eqref{e:bexp},~and~\eqref{e:hexp} to compute the corresponding parameters ${a_{\rm dyn} =\beta K a}$, ${b_{\rm dyn} = \beta K b}$, and ${h_{\rm dyn} = \beta h}$ that, along with the initial conditions $x_1(0)$ and $x_2(0)$, fully specify the underlying dynamics through Eq.~\eqref{eq:dynamic}. Furthermore, using Eqs.~\eqref{e:mcc-1fpA},~and~\eqref{e:mcc-1fpB} we can estimate the corresponding values for the slopes and intercepts describing the MCC rule and then compare with the experimental values reported. If the values obtained happen to be outside the credible interval ${[{O}^\ast-\zeta\delta O^\ast, {O}^\ast+\zeta\delta O^\ast]}$ defined by the choice of parameter $\zeta$, then such a specific value for the parameters $\Theta$ are rejected.

\subsection{\label{s:impl}Implementation}
We have used the package \texttt{pomp}~\cite{pomp} implemented in \textrm{R} to perform the Bayesian inference via particle Markov chain Monte Carlo with an adaptive random walk as proposal distribution. This is a package specifically designed for parameter inference of partially observed Markov processes. 
\subsection{\label{s:results}Results}
\begin{table}\centering
\begin{tabular}{l r r}
\hline
Parameter & Heterogeneous network & Square lattice \\
\hline
$\widehat{A}$ & $1.16$  & $1.413$       \\
$\widehat{x}_0$ & $-0.24$  & $0.0483$     \\
$\widehat{y}_0$ & $-0.71$   & $-0.4346$   \\
$\widehat{\alpha}$ & $0.263$  & $0.4417$\\
\hline
\end{tabular}
\begin{tabular}{rrrr}
  \hline
  \multicolumn{4}{c}{Covariance matrix (Heterogeneous network)} \\ 
  \hline
$9.3 \times 10^{-2}$ & $6.7 \times 10^{-2}$ & $5.9 \times 10^{-2}$ & $-2.0 \times 10^{-2}$ \\ 
 $6.7 \times 10^{-2}$ & $5.1 \times 10^{-2}$ & $4.7 \times 10^{-2}$ & $-1.5 \times 10^{-2}$ \\ 
 $5.9 \times 10^{-2}$ & $4.7 \times 10^{-2}$ & $4.4 \times 10^{-2}$ & $-1.3 \times 10^{-2}$ \\ 
 $-2.0 \times 10^{-2}$ & $-1.5 \times 10^{-2}$ & $-1.3 \times 10^{-2}$ & $4.4 \times 10^{-3}$ \\ 
   \hline
\end{tabular}
\begin{tabular}{rrrr}
  \hline
  \multicolumn{4}{c}{Covariance matrix (Square lattice)} \\ 
  \hline
 $1.6 \times 10^{-3}$ & $4.3 \times 10^{-5}$ & $-1.0 \times 10^{-4}$ & $-5.7 \times 10^{-5}$ \\ 
   $4.3 \times 10^{-5}$ & $6.8 \times 10^{-5}$ & $5.4 \times 10^{-5}$ & $1.6 \times 10^{-6}$ \\ 
   $-1.0 \times 10^{-4}$ & $5.4 \times 10^{-5}$ & $7.4 \times 10^{-5}$ & $1.2 \times 10^{-5}$ \\ 
   $-5.7 \times 10^{-5}$ & $1.6 \times 10^{-6}$ & $1.2 \times 10^{-5}$ & $1.3 \times 10^{-5}$ \\ 
   \hline
\end{tabular}
\caption{\label{t:diagramZH} Parameters locating on the phase diagram of the model (Fig.~\ref{fig:criticality} in the main text) the experiment performed in Zaragoza on a heterogeneous network and a square lattice~\cite{garcia-lazaroEtAl2012}. These data were obtained from Table~\ref{t:ZH} by using Eqs.~\eqref{e:alphaexp},~\eqref{e:Ainf},~\eqref{e:x0inf},~and~\eqref{e:y0inf} to get the parameter values and first order error propagation to get the corresponding covariance matrix.}
\end{table}

Table~\ref{t:data} summarizes the experimental results reported in Ref.~\cite{garcia-lazaroEtAl2012}. The quantity $\bar{x}$ represents the global level of cooperation reached by the system of interacting humans in the laboratory. We estimate this quantity and its standard error by computing the average and standard deviation, respectively, of the global cooperation [Figs.~\ref{fig:results}(a) and~\ref{fig:results}(b) in main text] over the last ten rounds of each of the two experiments performed in Zaragoza~\cite{garcia-lazaroEtAl2012}, on an heterogeneous network and on a square lattice. These are the two experiments that we analyze in this manuscript and to which we refer to in this section.

Table~\ref{t:ZH} shows the parameters estimated for the two experiments and their corresponding covariance matrix. In the case of the experiment on an heterogeneous network (respectively square lattice) we used $\zeta=1.96$ (respectively $\zeta=1.28$) corresponding to a uniform distribution on the experimental quantities representing a $97.5\%$ (respectively $90\%$) credible interval (Sec.~\ref{s:mono}). Since these results were obtained in the regime of mono-stability (Sec.~\ref{s:mono}), the parameters over which we performed Bayesian inference were ${\Theta_{\rm exp} = (m_C , m_D , r, G, x(0))}$. (In the regime of bi-stability we did not find satisfactory results.) The dynamical parameters are then determined through  Eqs.~\eqref{e:alphaexp},~\eqref{e:hexp0},~\eqref{e:aexp0},~and~\eqref{e:bexp0}. Figs.~\ref{fig:results}(a) and~\ref{fig:results}(b) in the main text show the dynamics corresponding to these parameters in the case of the experiment on an heterogeneous network and on a square lattice, respectively. On the other hand, Figs.~\ref{fig:results}(c) and~\ref{fig:results}(d) show the results of applying Eq.~\eqref{e:dut} in the main text using these parameter values. 

Table~\ref{t:diagramZH} shows the corresponding values of the parameters $\Theta_{\rm dyn} = (\theta_{\rm phase}, \alpha)$, where $\theta_{\rm phase}=(A,x_0,y_0)$ directly locate the system in the phase diagram of the model (Fig.~\ref{fig:criticality} in the main text) and, along with $\alpha$, completely determine the dynamics of the system through Eq.~\eqref{eq:dynamic}. The covariance matrix reported is obtained by first order error propagation of the results displayed in Table~\ref{t:ZH}. This was done to take into account that the constraints were enforced on $\Theta_{\rm exp}$ during the inference process and produced the best visual results of Fig.~\ref{fig:results}. We also tried to first transform the posterior over the parameters $\Theta_{\rm exp}$  into a posterior on the parameters $\Theta_{\rm dyn}$ to then compute the average value over the latter, but the results were less satisfactory. In any case, Fig.~\ref{fig:criticality} in the main text also show the population of parameters $\Theta_{\rm dyn}$ representing the corresponding posterior. We see that the values reported in Table~\ref{t:diagramZH}, which are also shown in Fig.~\ref{fig:criticality} on the main text, indeed appear to be representative of the population.

Finally, we estimated the Euclidean distance ${d(\theta_{\rm phase},\theta_c^\ast)}$ of the parameters $\theta_{\rm phase}$ to the closest point ${\theta_c^\ast = (A_c^\ast,x_{0c}^\ast,y_{0c}^\ast)}$ on the critical lines defined by Eqs.~\eqref{e:y0c},~\eqref{e:Ac},~and~\eqref{e:x0c}. Following the standard analysis of continuous phase transitions, we define a reduced or relative distance to the critical point as $\delta(\Theta_{\rm phase},\theta_c) = d(\theta_{\rm phase},\theta_c)/|\Theta_c|$, where $|\theta_c|$ stands for the Euclidean norm of the vector of parameters $\theta_c$. Using the values in Tab.~\ref{t:diagramZH} we obtained the values of $\delta(\Theta_{\rm phase},\Theta_c)\approx 0.03$ and $\delta(\Theta_{\rm phase},\Theta_c)\approx 0.11$ for the experiments on an heterogeneous network and on a square lattice, respectively. 

\let\oldaddcontentsline\addcontentsline
\renewcommand{\addcontentsline}[3]{}

\let\addcontentsline\oldaddcontentsline

\end{document}